\documentclass[11pt]{article}

\newcommand{\ud}{\textrm{d}}

\newcommand{\commut}[2]{[\,#1\,,\,#2\,]}

\usepackage{amsmath}
\usepackage{amssymb}
\usepackage{amsfonts}
\newcommand{\tr}{\textrm{tr}}
\newcommand{\Tr}{\textrm{Tr}}

\usepackage{graphicx}
\usepackage{epsfig}

\def\pslash{\hbox{/\kern-.5800em$p$}}

\def\gappeq{\mathrel{\rlap {\raise.5ex\hbox{$>$}}
{\lower.5ex\hbox{$\sim$}}}}
\def\lappeq{\mathrel{\rlap{\raise.5ex\hbox{$<$}}
{\lower.5ex\hbox{$\sim$}}}}
\newlength{\dummysp}
\newcommand{\beq}{\begin{eqnarray}}
\newcommand{\eeq}{\end{eqnarray}}

\newcommand{\ben}{\begin{enumerate}}
\newcommand{\een}{\end{enumerate}}

\newcommand{\bit}{\begin{itemize}}
\newcommand{\eit}{\end{itemize}}

\def\({\left (}
\def\){\right )}

\begin{document}

\pagestyle{empty}

\

\

\

\

\

\

 \begin{center}
\bigskip
 { \Large \bf Lattice Chirality and the Decoupling of Mirror Fermions }

\bigskip

\bigskip

\bigskip
 {\sc Erich Poppitz} and {\sc Yanwen Shang}

\bigskip

{\it \small Department of Physics\\ University of Toronto\\ Toronto, ON M5S~1A7, Canada}

\bigskip

{\it \small Emails: poppitz@physics.utoronto.ca, ywshang@physics.utoronto.ca }

\

\

\vspace{2cm}

{\bf Abstract:} 
\end{center}

{\flushleft{W}}e show, using exact lattice chirality, that  partition functions of lattice gauge theories with vectorlike fermion representations  can be split into ``light" and ``mirror" parts, such that the ``light" and ``mirror" representations are chiral. The splitting of the full partition function into ``light" and ``mirror"  is well defined  only if the two sectors are separately anomaly free. We show that only then is  the  generating functional, and hence the spectrum, of the mirror theory   a smooth function of the gauge field background. This explains how  ideas to use additional non-gauge, high-scale mirror-sector dynamics to decouple the mirror fermions without breaking the gauge symmetry---for example, in symmetric phases at strong mirror Yukawa coupling---are forced to respect the anomaly-free condition when combined with the  exact lattice chiral symmetry. 
Our results are also useful in explaining a paradox posed by a recent numerical study of the mirror-fermion spectrum in a toy would-be-anomalous two-dimensional theory.   In passing, we  prove some general properties of the partition functions of arbitrary chiral theories on the lattice that should be of interest for further studies in this field.

\vfill
\begin{flushleft}
\end{flushleft}
\eject
\pagestyle{empty}

\setcounter{page}{1}
\setcounter{footnote}{0}
\pagestyle{plain}

\section{Introduction and summary}
\label{intro}

\subsection{Motivation}
\label{motivation}
The study of strong-coupling chiral gauge dynamics is an outstanding problem of great interest, both on its own and for its possible relevance to phenomenology.  Whereas the standard model of elementary particle physics is a weakly coupled chiral gauge theory, additional strong chiral gauge dynamics at (multi-) TeV scales may be responsible for breaking the electroweak symmetry and fermion mass generation.  

Several different approaches  are currently available for the study of the strong-coupling behavior of chiral gauge theories.
Notably, one has 't Hooft's anomaly matching and most attractive channel arguments, which are
complemented by the ``power of holomorphy" in supersymmetric theories.  Scaling arguments and effective NJL-like models,
both using results from QCD as a stepping stone, have also been employed extensively; on the other hand, large-$N$ expansions, including the recently considered gravity
duals in the AdS/CFT (AdS/QCD) framework, do not usefully apply to
chiral gauge theories.  None of these approaches
represents a ``first principles'' method, with an
accuracy that can be systematically improved. The space-time lattice regularization
remains, to this day, the only way  offering hope for such a systematic progress.

During the past two decades, since the work 
of Ginsparg and Wilson (GW)  \cite{Ginsparg:1981bj}, there has been significant progress 
in understanding chiral symmetries on the 
lattice \cite{Kaplan:1992bt}--\cite{Luscher:1998pq}; further references 
are given in the reviews \cite{Golterman:2000hr, Luscher:2000hn}, 
while  \cite{Golterman:2004qv, Bhattacharya:2005xa, Bhattacharya:2006dc} contain more 
recent work.  

Recently,  the existence of an exactly gauge invariant\footnote{There also  exists a point of view  \cite{Neuberger:2001nb}  that an exact gauge invariance at finite lattice spacing may not be necessary and that ``gauge averaging" of the fermion determinant will wash out, by a mechanism due to  \cite{Fradkin:1978dv}, \cite{Forster:1980dg}, any   gauge-breaking effects in the continuum limit in the anomaly-free case. While some numerical evidence supports this view  \cite{Narayanan:1996kz}, \cite{Izubuchi:1999kk},  the issue is far from settled, see  \cite{iz2}  and the review \cite{Golterman:2000hr}. For other ideas giving up exact gauge invariance, see \cite{HS}, \cite{Golterman:2004qv}.}
 lattice  construction of anomaly-free chiral gauge theories using  exact lattice chiral symmetries 
 has been proven in several particular cases \cite{Luscher:2000hn, Luscher:1998du}. However, an explicit formulation of the action and measure 
 outside of perturbation theory is currently not available.  We thus believe that the further study of the problem, the consideration of new proposals, and of their relationship to old and new advances in the field is a worthwhile task. 

\subsection{What this paper is about }
\label{whatabout}

In essence, this paper is about revisiting an old idea 
  \cite{Eichten:1985ft} 
in light of the new understanding of exact lattice chirality. The idea  is to begin with a vectorlike theory, whose explicit lattice formulation poses no problems.
The chiral components  of the vectorlike fermions are  split into  ``light" and ``mirror" fermions. The ``light"  fermions have the  chirality and group representations of the desired target chiral gauge theory and the ``mirrors" are simply their opposite-chirality partners. The vectorlike theory is then  deformed   in a way that (ideally) only affects the ``mirror" fermions: for example, one adds appropriate Yukawa interactions, or four- and multi-fermion interactions. The goal of the deformation is  to ensure that, when the parameters of the deformation are chosen appropriately, the mirror fermions decouple without breaking the gauge symmetry. Thus, at low energies,   the desired unbroken chiral gauge theory is supposed to emerge.

The decoupling of ``mirror" fermions in chiral representations without breaking the chiral  symmetry (which is gauged in chiral gauge theories) is possible in  the so-called strong-coupling symmetric phases of lattice Yukawa  \cite{Hasenfratz:1988vc, Stephanov:1990pc, Golterman:1990zu, Golterman:1991re} or multi-fermion-interaction \cite{Eichten:1985ft}  theories (earlier, the possibility of nonzero fermion masses without chiral symmetry breaking has been  discussed, for  two dimensional models,  in \cite{Witten:1978qu}).

From the continuum physics point of view, the strong-coupling symmetric phases are a  lattice artifact. Their existence, in either two or four dimensions, is established   using the lattice strong-coupling expansion, where all correlations have a range smaller than the lattice spacing. One can thus say that in   models with a strong-coupling symmetric phase and heavy mirrors, their     mass is ``higher than the ultraviolet cutoff"---the physics at high scales being that of  lattice particles  with small  site-to-site hopping probability. The strong-coupling symmetric phases of lattice Yukawa or multi-fermion interaction models are thus analogous to the well-known high-temperature disordered (hence symmetric) phases of spin systems.

For the continuum physicist, who is unlikely to proceed past this Introduction, we will now give a cartoon-like continuum description of the physics. This will also serve to illustrate  the idea behind using strong interactions to decouple the mirrors and help us state the main issues we would like to address.

Consider thus the classic example  
 of a four-dimensional chiral gauge theory with nontrivial strong-coupling dynamics---see, e.g., \cite{Raby:1979my}---an $SU(5)$ gauge theory with a ${\bf 5^*}$ and a ${\bf 10}$ Weyl fermion representation. We use two-component spinor notation to describe the desired ``light" fermions:
\beq
\label{su5-1}
\psi^i_\alpha \sim {\bf 5^*} ,~~ \chi_{ij \; \alpha}  \sim {\bf 10}~,~~\zeta_\alpha \sim {\bf 1}~,
\eeq
(here  $i$ denotes an $SU(5)$ (anti-)fundamental and $\alpha = 1,2,$ an $SL(2,C)$ index) and their  ``mirror" partners: 
\beq
\label{su5-2}
\eta_{i \alpha} \sim {\bf 5} ,~~ \rho^{ij}_\alpha  \sim {\bf 10^*}~, ~~\xi_\alpha \sim {\bf 1}~.
\eeq
In this notation a Dirac mass term for the ${\bf 5}$ would be $\psi^{i \alpha} \eta_{i \alpha} + {\rm h.c.}$. The gauge singlet  Dirac fermion (with Weyl components  $\zeta, \xi$) is a field whose $\xi$ component will play  an important role in the strong mirror dynamics; an entire singlet Dirac multiplet was added to make sure the fermion representation (\ref{su5-1}, \ref{su5-2}) is vectorlike and thus easy to put on the lattice.

The target $SU(5)$ chiral gauge theory has one anomaly-free $U(1)$ global symmetry, under which $\psi^i$ has charge $-3$ and $\chi_{ij}$ charge $1$; on the other hand, the vectorlike theory with fermion content (\ref{su5-1}, \ref{su5-2}) has  more exact global symmetries. 
Now, to decouple the mirrors (\ref{su5-3}), one adds interactions involving (ideally) only the mirror fields, of the form:
\beq
\label{su5-3}
\lambda  \; \xi^{\alpha}\; \eta^i_\alpha  \eta^{j \; \beta}  \rho_{ij \; \beta} + \ldots \eeq 
where the dots denote terms needed to break the extra global symmetries. 

The main insight helping to  decouple  the mirrors  is the realization that   the strong  lattice four-fermi interaction (\ref{su5-3}) can lead  to the formation of $SU(5)$ invariant mirror composite states, which can  acquire mass without breaking $SU(5)$.  We stress again that the strong-coupling symmetric phase and the formation of the singlet mirror composite states requires $\lambda  \gg1 $ in UV-cutoff units; this only makes sense on the lattice, and  the spectrum can be studied using the strong-coupling expansion (for details, see the appendix of ref.~\cite{Eichten:1985ft}).  
For example, a possible composite of the mirror fermions is the $\eta\eta\rho$ (${\bf 5}$-${\bf 5}$-${\bf 10^*}$) invariant appearing in (\ref{su5-3}).  It can   acquire a large Dirac mass by pairing with the singlet mirror field $\xi$ and can thus decouple  from the low-energy physics without breaking the $SU(5)$ symmetry;  at strong coupling all mirror fermions are similarly bound in massive  composites and decouple.\footnote{Ideas involving strong-Yukawa symmetric phases work similarly  \cite{Hasenfratz:1988vc, Stephanov:1990pc, Golterman:1990zu, Golterman:1991re} and are closely  related to  the multi-fermion interaction ones \cite{Golterman:1992yh}.} Since   the $SU(5)$ gauge interaction   is asymptotically free,  the strong mirror dynamics at and above the cutoff scale should   have   a parametrically small effect on the infrared chiral dynamics. Thus, the desired unbroken chiral gauge theory with massless fermion spectrum (\ref{su5-1}) is recovered in the infrared.

The idea to decouple the mirrors in the way described above is attractive and many important advances in understanding the strong-coupling symmetric phases in both Yukawa and multi-fermion-interaction theories were made in the past. However, in all cases studied, the spectrum of massless  fermion  states  (when they existed) was found to be vectorlike:
\begin{enumerate}
\item
The most notable reason is  the fact that on the lattice---until the recent advances in  exact lattice chirality---there was no way to define chiral components of the spinors at finite lattice spacing while avoiding fermion doubling. Because chiral symmetries are broken on the lattice with the traditional Wilson formulation, the strong interactions   (\ref{su5-3}) were, in all cases, also ``felt" by the ``light" fermions, causing either the ``light" fermions  to obtain mass or the ``mirror" fermions to become massless;  see \cite{Golterman:1993th,  Golterman:1994at, Golterman:1992yh}.
\item
Moreover, since  the $SU(5)$ gauge dynamics is only a spectator of the strong interactions whose  purpose is to   decouple  the mirrors, it is not clear how the non-gauge strong interactions of the mirror sector were supposed to ``know" about   chiral gauge anomalies and how they would decide to ``enforce"  the anomaly-freedom requirement on the light chiral spectrum or else,  break the gauge symmetry 
\cite{Jackiw:1984zi, Halliday:1985tg, Neuberger:2000wq, Matsui:2004dc}.\footnote{For example, ref.~\cite{Golterman:1992yh} found that in the 
model of \cite{Eichten:1985ft} the  non-gauge mirror dynamics was essentially the same 
  in models with anomaly-free  and anomalous fermion spectrum.}
\end{enumerate}
In this paper, we will address the above issues by using the exact chirality-preserving GW-fermion formulation of lattice vectorlike theories via the Neuberger-Dirac operator:
\begin{itemize}
\item It is well known that the GW-fermion formulation allows one to define chiral components of the spinors at finite lattice spacing without introducing doublers. The ``light" chiral components can be then excluded from participating in the strong ``mirror" interactions, like the one in (\ref{su5-3}).  The lattice theory can then be arranged to have exactly the  global symmetries and anomalies of the target continuum chiral theory, something that earlier Yukawa or four-fermi proposals could not achieve   \cite{Bhattacharya:2006dc}.
\item Furthermore, by considering in  detail the split of the vectorlike lattice partition function into ``light" and ``mirror" parts  in an arbitrary gauge background,   we will show that the anomaly-free condition on the light spectrum is also enforced by consistency of   the GW formulation of lattice chirality. We will make extensive use of the work of Neuberger \cite{Neuberger:1998xn} and L\" uscher \cite{Luscher:1998du} on chiral anomalies in the overlap/GW-fermion  formalism, see also \cite{Adams:2000yi}
\end{itemize}
The proposal to use strong-coupling Yukawa models with GW fermions to decouple the mirrors in a vectorlike gauge theory was made in \cite{Bhattacharya:2006dc}, where the many desirable features of such a formulation were pointed out.\footnote{We note that ref.~\cite{Creutz:1996xc} made an earlier suggestion along similar lines,  in the framework of a  domain wall with a finite fifth dimension.}  The proposal is attractive, as it gives an explicit gauge-invariant definition of the measure and lattice action, and because it has all the right symmetries and  anomalies of the target chiral gauge theory already at finite lattice spacing. However, 
this elegance comes at a price---the study of the mirror dynamics at strong coupling, which is needed to show that the mirrors do indeed decouple, is complicated by the exponential-only locality of the Neuberger-Dirac operator \cite{Hernandez:1998et, Neuberger:1999pz}. The strong-coupling expansions used  in relatively simple models \cite{Eichten:1985ft, Hasenfratz:1988vc, Stephanov:1990pc, Golterman:1990zu, Golterman:1991re, Golterman:1993th,  Golterman:1994at, Golterman:1992yh} to predict the formation of heavy fermion composites without breaking the chiral symmetry  are not easy to implement and a Monte-Carlo study is called for. 

A numerical study of the  strong-coupling mirror dynamics of a toy two-dimesional model  was performed in \cite{Giedt:2007qg}, for  a vanishing gauge background. The numerical  evidence found there indicates that, indeed, the mirror sector decouples at strong mirror Yukawa coupling. The questions of gauge anomalies in  the light target theory  and the ways the   dynamics would prevent them was not addressed. This is the main issue we focus on in this paper.

\subsection{Outline  and summary of results}
\label{outline}

Much of the discussion in this paper is rather technical. Here we outline the main points and summarize our results.  The reader is assumed to be familiar with the GW relation and the exact lattice chiral symmetry in vectorlike theories; for a review, see \cite{Luscher:2000hn} and references therein (our notation for the Neuberger-Dirac operator, the modified-$\gamma_5$ projectors,  and their eigenvectors is established in Section \ref{gen1}).

In Section \ref{gen2}, we consider in detail how the   partition function of a vectorlike theory splits into left- and right-  chirality components, using the eigenvectors of the modified-$\gamma_5$ as basis. We also explicitly work out the transformations of the left- and right- chirality partition functions and of the Jacobian under changes of the gauge background. 

In Section \ref{paradox}, we turn to the description of what we call the ``1-0" model: a toy two dimensional model, used   in a Monte-Carlo study of the decoupling of the mirrors in the strong-Yukawa symmetric phase \cite{Giedt:2007qg}. We  show how the partition function of this model (with vectorlike fermion content)  splits into a ``light" and ``mirror" part in an arbitrary gauge background. Only the ``mirror" degrees of freedom participate in the strong Yukawa interaction, which is introduced to decouple the ``mirrors" from the long-distance physics  (similar in spirit to (\ref{su5-3})). 

Using the results of Section \ref{gen2}, we then work out the gauge transformations of the ``light" and ``mirror" partition functions and show that the gauge transformation of the ``mirror" partition function precisely cancels the anomaly of the light fermions, independent of the value of the mirror Yukawa coupling(s) and for arbitrary gauge backgrounds.

We then contrast this finding with 
 the numerical results of 
\cite{Giedt:2007qg}. The Monte-Carlo simulation of the mirror dynamics at strong Yukawa coupling and in vanishing gauge background  provided  evidence for the  decoupling of the mirror sector without breaking the gauge symmetry (i.e., of the  existence of the desired strong-Yukawa symmetric phase with heavy mirrors). The massless spectrum of the theory consists of a left-handed fermion of unit charge  and a right-handed singlet under the gauge group.
These numerical results, combined with the exact gauge transforms of the mirror  partition functions worked out above, present us with a paradox. If the  decoupling of the mirrors at strong Yukawa coupling and zero gauge background persists also for an infinitesimal gauge background, as one would naively expect based on ``continuity," it is not clear how the heavy degrees of freedom could conspire to cancel the anomaly of the massless fermions.
(See also the Addendum for more discussion.) 

As already alluded to, the resolution of the paradox is in the assumption of the continuity. It turns out that the ``light"--``mirror" split of the partition function of the vectorlike theory is only well-defined if the light and mirror representation are separately anomaly-free. The results of the Monte-Carlo simulations in the ``1-0" model  hold for the trivial $(U=1)$ gauge background. However,   we will show that the mirror partition function is not a smooth function of the gauge background precisely at $U=1$ and that this singularity  prevents any  discussion of the mirror spectrum in general backgrounds.

We explain in detail how this comes about in Sections \ref{choosingvectors} and \ref{nirvana}.   There, while still with the ``1-0" model in mind, we switch the focus of our discussion to the most general form of chiral theories and study their properties based only on their defining characteristics.  Many results found there are therefore of general applicability and should be important for further studies of chiral theories on the lattice.

We begin, in Section \ref{changeofvectors} with a discussion of the dependence of the modified-chirality basis vectors on the gauge-field background; the material of this section is known, but for completeness we present a short self-contained derivation.

 In Section \ref{Yanwen'sTheorem},  we prove our main result on the variation of the most general chiral partition function under changes of the gauge background. We show that the variation of the partition function defined with an arbitrary chiral action always factorizes into a variation that depends on the basis vectors and a variation only due to the dependence of the operators included in the action on the gauge background.  This generalizes the known results  for simple actions;  see  Section (\ref{gen2}).  It is an important piece of knowledge since it isolates the anomalies from the details of the chiral theory and manifestly realizes on the lattice the idea that anomalies  are determined only by the representation  of the fields and not by the details of the Lagrangian. It has at least a few surprisingly powerful implications, one of which will be explained in Section~\ref{nirvana}.

In Section \ref{anomalywilson} and \ref{measure} we explain how, in the case of an anomalous representation, the chiral fermion measure can not be defined as a smooth function of the gauge background. We use the Wilson-line subspace of the gauge field background to illustrate, following \cite{Neuberger:1998xn},  the topological obstruction of defining a smooth fermion measure due to the anomaly. 
We explicitly show that in our ``1-0" toy model the mirror-fermion measure is not a smooth function of the gauge fields exactly at vanishing gauge background. We then explicitly demonstrate (within the Wilson-line subspace only) how to construct smooth fermion measures in the case of anomaly-free representations, for example, in the ``3-4-5" model \cite{Bhattacharya:2006dc} by showing how the singularities due to different representations can cancel each other  and how the phase ambiguity of the chiral partition function enters to help.

Finally, in Section \ref{nirvana}, we consider an interesting application of the results proven in Section \ref{Yanwen'sTheorem}  to show that the generating functional of the mirror theory is indeed a smooth function of the gauge background as long as the mirror representation is anomaly free. Thus, as an encouraging message of this paper, one expects that a demonstration of the decoupling of the mirror sector in an anomaly free model at vanishing gauge background will persist, by smoothness, also for small gauge background, e.g., in perturbation theory with respect to the gauge coupling.  The proof given in this section is also a general result independent on the details of the mirror action, and therefore the conclusion found there remains true for  any well-behaved chiral theory as long as the anomaly-free condition is satisfied.

\subsection{Outlook}
\label{outlook}

The main statement we make in this paper is that the splitting of vectorlike partition functions into ``light" and ``mirror" parts is smooth, as a function of the gauge field background, only if the mirror and light fermion representations are separately anomaly free. This is a comforting result as it explains how the non-gauge dynamics introduced to decouple the mirrors is forced to obey the anomaly free condition, if one wishes to generalize the results to a full theory with dynamical gauge fields. If the gauge field is taken as fixed external background only, there might also exist other mechanisms that force the anomaly cancellation conditions as explained in the Addendum.

We think that this result encourages further study of the decoupling of the mirror fermions in anomaly-free representations via strong lattice-cutoff-scale dynamics, such as that of strong-Yukawa symmetric phases. The next most important question is, of course, to demonstrate that the strong mirror dynamics does indeed cause the mirrors to decouple in anomaly-free cases and for trivial gauge background. 

We stress the main   advantages of the approach studied here: the fermion measure is well defined (as it is the trivial measure of the vectorlike theory), the global symmetries are realized exactly as in the desired target theory, and the partition function is exactly gauge invariant. Symmetry and beauty aside, the ultimate goal  of the approach 
is to be useful for actual numerical simulations of chiral gauge theories. Whether this will happen depends on many yet unknown factors, notably the possible complexity or sign problem of the  partition function. Here we only note that, in zero gauge background, the partition function of the ``1-0" model at infinite Yukawa coupling was found in \cite{Giedt:2007qg}  to be real and positive; this raises hopes that in the theory with dynamical gauge fields  the phase problem may be not too severe  at large values of the Yukawa coupling. This issue certainly deserves more attention. 

Finally, as already mentioned, the analytic strong-coupling expansion using the Neuberger-Dirac operator is complicated by its exponential-only locality \cite{Hernandez:1998et, Neuberger:1999pz}, leading one to suspect that Monte-Carlo simulations may appear as the only tool to study the strong-coupling mirror dynamics. However, 
we  note the recent work \cite{Gerhold:2006rc} on an analytic strong-coupling expansion in some four-dimensional Yukawa models with GW fermions (at vanishing gauge background). Within the approximations used, analytic evidence---backed up by results of recent Monte-Carlo simulations \cite{Gerhold1}---for the existence of a strong-coupling symmetric phase was found. It may thus be interesting to study the possible application of 
these methods to models designed to decouple mirror fermions.

\section{Splitting  partition functions of vectorlike theories into chiral components}
\label{gen}

\subsection{Notations and basis vectors}
\label{gen1}

{\flushleft{I}}n terms of the massive Wilson operator $D_W$,  the  modified-$\gamma_5$ matrix $\hat\gamma_5$ and the Neuberger-Dirac operator $D$ are  expressed as \cite{Luscher:2000hn}: 
\beq
\label{hatgamma5}
\hat\gamma_5 &=& {\gamma_5 A \over \sqrt{ (\gamma_5 A)^2}}, ~~ A \equiv 1 - D_W , ~~ D \equiv  1 - \gamma_5 \hat\gamma_5~,
\eeq
where $D$ transforms covariantly under gauge transforms, $D_{xy} \rightarrow e^{i \omega_x} D_{xy} e^{- i \omega_y}$ and the Ginsparg-Wilson (GW) relation is equivalent to $\hat\gamma_5^2 = 1$. 
Next define the following complete set of states:
\beq
\label{uwbasis}
\hat\gamma_5 u_i &=& - u_i ~~,~~ ~~~~~\hat\gamma_5 w_i  = w_i \\
\hat{P}_- &=& \sum_i u_i u_i^\dagger ~~,~~ \hat{P}_+  =  \sum_i w_i w_i^\dagger = 1 - \hat{P}_- ~,
\eeq
where we treat $u,w$ as columns and $u^\dagger, w^\dagger$ as rows.
For a topologically trivial background, the number of $u$ and $w$ eigenvectors is the same, equal to $N^2$ each for a two-dimensional square lattice ($2 N^2$ total dimension).\footnote{Most of the formulae in this paper are valid in any even dimension; in a few obvious instances, however, we specialize to two dimensions. Also, when necessary, we specialize to the case of a $U(1)$ gauge group.}  We also define the eigenvectors of $\gamma_5$, which can be chosen independent of the gauge background:
\beq
\label{vtbasis}
\gamma_5 v_i &=& v_i ~~,~~~~~~~ \gamma_5 t_i = - t_i \\
 {P}_+ &=& \sum\limits_i v_i v_i^\dagger ~~,~~ {P}_-  = \sum\limits_i t_i t_i^\dagger = 1 -  P_+ ~.
\eeq

 \subsection{Chiral variables, Jacobians, and their variations}
 \label{gen2}
 
 Consider a  vectorlike lattice theory with partition function:
 \beq
 \label{z0}
 Z_V &=& \int \prod_x \ud \Psi_x \ud \bar\Psi_x \; e^{S}~,
 \eeq
 where $x$ denotes both spinor and spacetime lattice indices.
For the time being, we will take the action $S$ to be the  usual kinetic action $S = \sum_{x, y} \bar\psi_x D_{x, y} \psi_y \equiv (\bar\Psi \cdot D \cdot \Psi)$, 
which has an exact chiral symmetry, $\Psi \rightarrow e^{i \alpha \hat\gamma_5} \Psi$, $\bar\Psi \rightarrow \bar\Psi e^{i \alpha \gamma_5}$.
 
Now we   change  variables from $\Psi_x$, $\bar\Psi_x$ to $c_i^{\pm}$, $\bar{c}_i^\pm$ defined by the following expansions in terms of the $\gamma_5$ and $\hat\gamma_5$ eigenvectors (we let $x$ also include spinor index, thus $x$ takes $2N^2$ values in 2d):
\beq
\label{psitoc}
\Psi_x &=& \sum_i c_i^+ w_i (x) +  c_i^- u_i(x) \\
\bar\Psi_x &=& \sum_i \bar{c}_i^+ t_i^\dagger (x) +  \bar{c}_i^- v_i^\dagger(x)~.
\eeq
The change of variables 
leads to a Jacobian: 
\beq
\label{J1}
\prod_x \ud \Psi_x \ud \bar\Psi_x &=& {1 \over  J}\;  \prod_i \ud c_i^+ \ud c_i^- \ud \bar{c}^+_i \ud \bar{c}^-_i  \;\\
J&=& 
{\rm det} || w_i(x) u_j(x) ||\;{\rm det} || v_i^\dagger (x) t_j^\dagger (x) ||~,
\eeq
(note that $|| w_i(x) u_j(x) ||$ is a $2N^2 \times 2N^2$ dimensional matrix, with $x$ indexing rows and $i,j$-columns)
and the partition function becomes:
\beq
\label{z1}
Z_V &=& \int \prod_x \ud \Psi_x \ud \bar\Psi_x e^{S} = {1 \over  J}\;\int  \prod_i \ud c_i^+ \ud c_i^- \ud \bar{c}^+_i \ud \bar{c}^-_i
e^{\sum_{i,j} \bar{c}^+_i c^+_j (t_i^\dagger \cdot D \cdot w_j) + \bar{c}^-_i c^-_j (v_i^\dagger \cdot D \cdot u_j)} \nonumber \\
&=& {1 \over  J}\;   {\rm det} || (t_i^\dagger \cdot D \cdot w_j)|| ~ {\rm det} || (v_i^\dagger \cdot D \cdot u_j)||~.
\eeq
Under infinitesimal changes of the gauge field background:
\beq 
\label{changes}
U_{x, \mu}  \rightarrow   U_{x, \mu} + \delta_{\eta_{x, \mu}} U_{x, \mu}~ ,
\eeq
which, in the case of  gauge transformations,  take the form:
\beq
\label{gaugevariation}
 \delta_\omega U_{x,\mu} \big\vert_{gauge} = i \left(  \omega_x U_{x,\mu} - U_{x, \mu} \omega_{x + \mu} \right) \equiv - i \left( \nabla_\mu \omega_x \right) \; U_{x, \mu}~,
\eeq
the various factors in $Z_V$ change as described below.
\begin{enumerate}
\item The change of the ``positive chirality" determinant is:
\beq
\label{deltaMplus}
 &&\delta_\eta \ln\det || (t_i^\dagger \cdot D \cdot w_j)|| \\
&=& \sum_{j,k} (w_j^\dagger \cdot D^{-1} \cdot t_k )\; ( t_k^\dagger \cdot \delta_\eta D \cdot w_j) + (w_j^\dagger \cdot D^{-1} \cdot t_k )\; ( t_k^\dagger \cdot D \cdot \delta_\eta w_j) \nonumber\\
&=& \tr(  \hat{P}_+ D^{-1} \delta_\eta D) + \sum_j (w_j^\dagger \cdot \delta_\eta w_j)~. \nonumber
\eeq
To obtain (\ref{deltaMplus}), in the first line  we used  $\sum_k (w_j^\dagger \cdot D^{-1} \cdot t_k ) ( t_k^\dagger \cdot  D \cdot w_i) =\delta_{ji}$, while in the second line we used the freedom to insert  $\sum_k v_k v_k^\dagger$ (which, using $\hat{P}_+ D^{-1} = D^{-1} P_-$, is killed by the projectors); finally, we used completeness, $\sum_k t_k t_k^\dagger + v_k v_k^\dagger = 1$. The trace in (\ref{deltaMplus}) is over spinor as well as space-time indices.

We note that the first term in (\ref{deltaMplus}) reflects the change of the operator, $D$, while the second is due to the change of basis vectors $w_i$, which depend on the gauge background (while the $t, v$-vectors do not).  We stress that this factorization of the  change of the ``positive chirality" determinant into separate terms, one due to the change of the operators and the other due to the change of basis vectors,  is a general feature of chiral partition functions. This will be proven for partition functions defined with a general chiral action in  Section \ref{Yanwen'sTheorem}, and will be important in what follows.

\item For the ``negative chirality" determinant, using  
$\sum_k  (u_j^\dagger \cdot D^{-1} \cdot v_k ) ( v_k^\dagger \cdot   D \cdot u_i) = \delta_{ji}$, similar to 
the derivation of (\ref{deltaMplus}), we find:
\beq
\label{deltaMminus}
&&\delta_\eta \ln\det || (v_i^\dagger \cdot D \cdot u_j)||  \\
&=& \sum_{j,k} (u_j^\dagger \cdot D^{-1} \cdot v_k )\; ( v_k^\dagger \cdot \delta_\eta D \cdot u_j) + (u_j^\dagger \cdot D^{-1} \cdot v_k )\; ( v_k^\dagger \cdot D \cdot \delta_\eta u_j) , \nonumber \\
&=& \tr(  \hat{P}_- D^{-1} \delta_\eta D) + \sum_j (u_j^\dagger \cdot \delta_\eta u_j)~.\nonumber
\eeq
Here, we also have a contribution from the change of operator, the first term in (\ref{deltaMminus}) as well as a contribution due to the change of basis. 

\item Finally, the change of Jacobian is computed from
the change of its first factor: 
\beq
\label{deltaJ1}
\delta_\eta \ln \det || w_i(x) u_j(x) || &=&\sum_{x,y,i,j} \big\| \begin{array}{c} w^\dagger_i (x) \cr u^\dagger_j (y)\end{array} \big\| \times  || \delta_\eta w_i (x) \delta_\eta u_j(y) || \nonumber \\
&=& \sum_i (w^\dagger_i \cdot \delta_\eta w_i) + (u^\dagger_i  \cdot \delta_\eta u_i)~,
\eeq
leading to:
\beq
\label{deltaJ}
{1 \over J} \rightarrow {1 \over J} \; e^{ - \sum_i \left[ (w^\dagger_i \cdot \delta_\eta w_i) + (u^\dagger_i  \cdot \delta_\eta u_i)
 \right]  }~.
\eeq
\end{enumerate}
Now we can collect all factors, and find that in the vectorlike theory the factors in (\ref{deltaMplus}, \ref{deltaMminus}, \ref{deltaJ})  having to do with the choice of basis vectors cancel out from the change of the partition function and we are left with: 
\beq
\label{deltaZ}
Z_V[U + \delta_\eta U] &=& Z_V[U] e^{ \tr(  \hat{P}_+ D^{-1} \delta_\eta D)  +  \tr(  \hat{P}_- D^{-1} \delta_\eta D)  }  \nonumber \\
&=& Z_V[U]  e^{ \tr  D^{-1} \delta_\eta D }~,
\eeq
showing that  the change of the partition function is determined solely by the change of the GW operator.
 In particular, for a gauge variation of $U$, eqn.~(\ref{gaugevariation}), we find immediately from (\ref{deltaZ}) that $Z_V[U+\delta_\omega U] = Z_V[U]$,  and also that:
 \beq
 \label{gaugevariation1}
  \tr  \hat{P}_+ D^{-1} \delta_\omega D  &=& i \tr (P_-  - \hat{P}_+)\omega =  \;- {i \over 2} \;\tr \omega \hat\gamma_5=  -{i \over 2}\; \sum_x\;  \omega_x \; {\rm tr} (\hat\gamma_5)_{xx} ~, \nonumber \\
   \tr  \hat{P}_- D^{-1} \delta_\omega D  &=& i \tr (P_+  - \hat{P}_-)\omega = \; {i \over 2}\;  \tr \omega \hat\gamma_5 =  {i \over 2}\; \sum_x\;  \omega_x \; {\rm tr} (\hat\gamma_5)_{xx} ~,   \eeq
 where the trace in the last line is over spinor indices only.
 The field tr$( \hat\gamma_5)_{xx}$ appearing in the  basis-independent gauge variations   (\ref{gaugevariation1}), is known to be a topological lattice field, which   expresses  the chiral anomaly on the lattice (this follows, e.g., from the index theorem of \cite{Hasenfratz:1998ri}, see also  \cite{Luscher:1998du, Fujikawa:1998if, Luscher:1999un}). Naturally, eqns.~(\ref{gaugevariation1}) show that the anomalies due to the left- and right-moving fermions cancel. 
 
Since we will be interested in splitting vectorlike theories' lattice partition functions with more general actions into chiral components, and in the dependence of these chiral components on the gauge field background, we will focus our discussion on the term $\sum_i (w_i^\dagger \cdot \delta_\eta w_i)$ (or similarly $\sum_i u_i^\dag \cdot \delta_\eta u_i$) in the following sections and pay great attention to the variations of the basis vectors with respect to changes of the gauge background. Following \cite{Luscher:1998du}, we will refer to $\sum_i (w_i^\dagger \cdot \delta_\eta w_i)$, and similar for $w_i \rightarrow u_i$, as ``measure terms'' since not only they depend on but also uniquely determine the fermion measure \cite{Luscher:1999un}.

\section{The ``1-0" GW-Yukawa model and a paradox}
\label{paradox}

The Yukawa-Higgs-GW-fermion model being considered here, which we call the ``1-0" model, is a $U(1)$ two-dimensional lattice gauge theory with one charged Dirac fermion $\psi$ of charge 1 and a neutral spectator  Dirac fermion $\chi$. 

Considering this theory is motivated by its simplicity: it is the minimal Higgs-Yukawa-GW-fermion model in two dimensions which holds the promise to yield, at strong Yukawa coupling, a chiral spectrum of charged fermions and is, at the same time, amenable to numerical simulations not requiring the use of extensive computing resources. The fermion part of the action of the ``1-0" model is:
\begin{eqnarray}
\label{toymodel}
S&=& S_{light} + S_{mirror} \\
S_{light} &=&
 \left( \bar\psi_+ \cdot D_1  \cdot  \psi_+\right) + \left( \bar\chi_- \cdot  D_0  \cdot \chi_-\right) \nonumber  \\
 S_{mirror} &=& \left( \bar\psi_- \cdot D_1 \cdot  \psi_-\right) + \left( \bar\chi_+ \cdot D_0 \cdot  \chi_+\right)  \nonumber \\
&+& y \left\{ \left( \bar\psi_-  \cdot \phi^*  \cdot \chi_+ \right) + \left( \bar\chi_+ \cdot  \phi \cdot  \psi_- \right)   + h \left[ \left( \psi_-^T \cdot \phi \gamma_2 \cdot  \chi_+ \right) - \left( \bar\chi_+ \cdot  \gamma_2 \cdot  \phi^* \cdot  \bar\psi_-^T \right) \right] \right\} \nonumber~.
\end{eqnarray}
 The chirality components for the charged and neutral fermions are defined, by projectors including  the appropriate Neuberger-Dirac operators (charged $D_1$ and neutral $D_0$) for the unbarred components, i.e. $ \psi_\pm = (1 \pm \hat{\gamma}_5)\psi/2$ .
The field $\phi_x = e^{i \eta_x}$, $|\eta_x| \le \pi$, is a unitary higgs field of unit charge with the usual kinetic term:
\begin{eqnarray}
\label{Skappa}
S_\kappa =  \frac{\kappa}{2}\; \sum_{x} \sum\limits_{\hat{\mu}} \left[ 2 - \left(\; \phi_x^* \; U_{x, x+ \hat\mu} \; \phi_{x+\hat\mu} + {\rm h.c.}\; \right) \right]~.
\end{eqnarray}
The inclusion of both Majorana and Dirac gauge invariant Yukawa terms is necessitated by the requirement that all global symmetries not present in the desired target chiral gauge theory be explicitly broken, see \cite{Golterman:2002ns},  \cite{Bhattacharya:2006dc}. Moreover, consistent with the symmetries, if the Majorana coupling $h$ vanishes, there are exact mirror-fermion zero modes  for arbitrary backgrounds $\phi_x$, which can not be lifted in the disordered phase  \cite{Giedt:2007qg}.

 From now on, we will call the fermion fields that participate in the Yukawa interactions the ``mirror" fields---these are the negative chirality component, $\psi_-$, of the charged $\psi$, and the positive chirality component, $\chi_+$, of the neutral $\chi$, while the fields $\psi_+$ and $\chi_-$ will be termed ``light."

The lattice action (\ref{toymodel}) completely defines the theory via a path integral over the charged and neutral fermion fields, the unitary higgs field, as well as the gauge fields. We will not consider the integral over the lattice gauge fields, but will study in detail the variation of the partition function with respect to the gauge background.

Our interest is in the symmetric phase of the unitary higgs theory
(expected to occur at $\kappa < \kappa_c \simeq 1$), where the higgs field acts---modulo correlations induced by $\kappa \ne 0$ and by  fermion backreaction---essentially as a random variable. Based on experience with strong-Yukawa expansions in theories with naive or Wilson fermions, it is expected  that in the large-$y$, fixed-$h$ limit, there is a symmetric phase where the fermions $\psi_-$ and $\chi_+$  decouple from the long distance physics. In the symmetric phase, this decoupling occurs without breaking the chiral symmetry, essentially by forming chiral-neutral composites of the fermions   with the scalar $\phi$, as  described around eqn.~(\ref{su5-3}) of Section \ref{whatabout}.

The  expected spectrum of light fields in the target theory consists of  the charged $\psi_+$ and the neutral $\chi_-$. The spectrum of the mirror theory was investigated numerically in \cite{Giedt:2007qg}, for vanishing gauge field background and in the infinite-$y$ limit. The evidence found there points towards decoupling of the mirror sector, with no breaking of the chiral symmetry of the mirror sector (this symmetry is gauged by the $U(1)$ gauge field).
 The analysis of ref.~\cite{Giedt:2007qg} was performed by first using the analogue of the formulae from Sections \ref{gen1}, \ref{gen2},
 for the case of vanishing gauge background. The eigenvectors of $\hat\gamma_5$ were explicitly worked out  and the splitting of the partition function into ``light" and ``mirror" was made manifest. Subsequently, a Monte Carlo simulation of the mirror sector in the infinite-$y$ limit was performed, yielding the above-cited results about the decoupling of all mirror sector fields (decoupling at infinite $y$ further requires $h>1$). However, a complete decoupling of the mirror is subtle. As discussed in the Addendum, we suspect that light degrees of freedom might still exist in a very contrived way, and further studies are required to clarify this dynamical issue; needless to say, this is under current investigation.
 
 As we showed in Section \ref{gen2}, 
the lattice fermion action (\ref{toymodel}) and the corresponding partition function easily split into light and mirror parts also in an arbitrary fixed gauge background. Only the charged eigenvectors (of both light and mirror fields) depend on the   background. By analogy with (\ref{z1}) we have a split of the partition function:
  \beq
  \label{z01}
  Z[U; y, h] = Z_L[U] \times {1 \over J[U]} \times Z_M[U; y, h]~.
  \eeq
Here $Z_L[U] = \det || (t_i^\dagger \cdot D \cdot w_j) || \times$(determinant  of neutral light spectator) is the light sector partition function. The jacobian $J$ is the product of the jacobians (\ref{J1}) for the charged and neutral sectors. Finally,  $Z_M$ denotes the mirror partition function---an integral over the charged mirrors, neutral mirrors, and unitary higgs field. The mirror fermion integral  is a determinant which includes a kinetic term, as in (\ref{z1}), but now also the Yukawa terms from (\ref{toymodel}), and is also averaged   over the random $\phi_x$ (we take $\kappa \rightarrow 0$). 
 
 Now, because the l.h.s. of (\ref{z01}) is manifestly gauge invariant, so is the r.h.s., since it is obtained  from the l.h.s. simply via a change of variables. 
 But we know how two of the factors on the r.h.s. transform under gauge transformations: the light partition function  $Z_L[U]$ and the Jacobian $1/J[U]$, for which
 we have, from (\ref{deltaMplus}), using  (\ref{gaugevariation1}):
 \beq
 \label{z02}
 {Z_L[U^\omega] \over J[U^\omega]}& \simeq & {Z_L[U] \over J[U]}  \exp\left(  - {i \over 2} \tr \omega \hat\gamma_5 - \sum_i ( u_i^\dagger \cdot \delta_\omega u_i) \right) . \eeq
  Therefore, from (\ref{z01}) and the fact that the l.h.s. is gauge invariant, it follows that the mirror partition function transforms, under gauge transformations, as follows:
  \beq
  \label{z03}
  Z_M[U^\omega; y, h] \simeq Z_M[U; y,h]  \exp\left(   {i \over 2} \tr \omega \hat\gamma_5 + \sum_i ( u_i^\dagger \cdot \delta_\omega u_i)  \right)   ,
  \eeq
  independent  not only on the values of the Yukawa couplings ($y, h$) but also most of the details of the mirror action; we note that our more general considerations of Section \ref{Yanwen'sTheorem} give a direct proof of this result. In passing, we  stress that we can not similarly infer the change of $Z_M[U; y,h]$ under arbitrary (i.e., not gauge transformations) changes of background, since we expect that the change of $Z[U; y,h]$ on the l.h.s. of 
 (\ref{z01}) under arbitrary variations of $U$ depends on $y,h$.
 
The gauge variation of the mirror partition function of eqn.~(\ref{z03}) leads us to a paradox.\footnote{We thank N. Arkani-Hamed, M. Golterman, B. Holdom, and Y. Shamir for asking pertinent questions about the anomaly.}   The exact result (\ref{z03}) shows that the gauge transformation of the mirror partition function should precisely cancel that of the light chiral fermion. If the mirror sector only involves heavy degrees of freedom, as the numerical results of \cite{Giedt:2007qg} suggest, and if these zero-background results persist for arbitrarily small gauge backgrounds (as one is inclined to expect), then 
 the mirror partition function should be  a local functional of the gauge background. By (\ref{z03}), this local functional's gauge variation must precisely cancel the anomaly of the light chiral fermion. However, this is known to be impossible, as the anomaly is not the variation of a local functional.
  
In what follows we will argue that this paradox has a natural resolution in the case when dynamical gauge fields are turned on , which can be found using the results of \cite{Neuberger:1998xn} and \cite{Luscher:1998du}. We will show that the paradox is (naturally) absent if the anomalies in the light and mirror sectors cancel separately. Moreover,  we will argue, in Section \ref{nirvana}, that the mirror partition function  and, more generally, the generating functional for connected correlation functions in the mirror sector, are smooth functions of the gauge field background in the anomaly-free-mirror case only.

 \section{More on the choice of basis vectors}
 
 \label{choosingvectors}
 
To explain the resolution mentioned above and the other results alluded to in the last paragraph (to be discussed in  Section \ref{nirvana}), we need to first consider in more detail the variations of the chiral basis vectors under arbitrary changes of the gauge background and the properties of the resulting fermion measure. This is important, since, as we saw in Section \ref{gen2} and will show for more general chiral partition functions in Section \ref{Yanwen'sTheorem}, the ``measure terms" determine the basis-vector-dependent part of the chiral partition functions' variation with the gauge background.  They reflect the ambiguity in the phase choice of the chiral partition functions.  Interestingly, the ``curvature", associated to this term thought of as a connection, is basis independent.  Therefore the ``measure term'' can not be chosen at random.  In particular, in the anomalous case (Section \ref{anomalywilson}) we recall why there is no definition of the ``measure terms" which is a smooth function of the gauge field background. We also explicitly show the singularity of the basis vectors and the associated ``measure terms" that were chosen in our analysis of the 1-0 model.  In the anomaly-free case (Section \ref{measure}) we show  how to construct  a smooth measure in the Wilson-line subspace of gauge field space by cancelling the singularities in the measure precisely with the help of the phase ambiguity.

    \subsection{Change of basis vectors under arbitrary change of background}
   \label{changeofvectors}
   
   The change of chiral partition function under arbitrary changes of the gauge background is of great interest. By eqns.~(\ref{deltaMplus}), (\ref{deltaMminus}) and also Section \ref{Yanwen'sTheorem}, this clearly depends on the change of basis vectors. Hence, we begin this Section by studying how the $\hat\gamma_5$ eigenvectors change under changes of the gauge background.
   
In a $U(1)$ gauge theory, a gauge (\ref{gaugevariation}) and an arbitrary (\ref{changes}) change of background differ in the choice of the function $\eta$ in (\ref{changes}):
  \beq
  \label{changes2}
  \delta_{\eta_{x, \mu}} U_{x, \mu} \equiv \eta_{x, \mu} U_{x, \mu} 
  \eeq
  where for gauge variations we have $\eta_{x, \mu} = - i \nabla_\mu \omega_x$. The $\hat\gamma_5$ matrix changes as follows:
  \beq
  \label{changeofgamma5}
  \hat{\gamma_5} [U + \delta_\eta U] = \hat\gamma_5[U] - \gamma_5 \delta_\eta D~,
  \eeq
  where the second term follows from $\hat\gamma_5 = \gamma_5(1- D)$. The variation $\delta_\eta D$ obeys:
  \beq
  \label{deltaD1}
  \hat\gamma_5 (\gamma_5 \delta_\eta D) = - (\gamma_5 \delta_\eta D) \hat\gamma_5~,
  \eeq
as a consequence of the GW relation (i.e., $\hat\gamma_5^2 = 1$).
  
  Now given the set of eigenvectors $u_i, w_i$ of $\hat\gamma_5[U]$, obeying orthonormality $(w_i^\dagger \cdot w_j) = (u_i^\dagger \cdot u_j) = \delta_{ij}$ and $(u_i^\dagger \cdot w_j) = 0$, we wish to find the eigenvectors of $\hat{\gamma_5} [U + \delta_\eta U]$ of (\ref{changeofgamma5}):
  \beq
  \label{newvectors}
  (\hat\gamma_5 - \gamma_5 \delta_\eta D) \;w^\prime_i &=& w^\prime_i \nonumber \\
  (\hat\gamma_5 - \gamma_5 \delta_\eta D) \; u^\prime_i &=& -u^\prime_i ~.
  \eeq
  We assume that in the neighborhood of the chosen initial background  $U$  the vectors change smoothly under small changes of the gauge background. We thus look for $w^{\prime}_i$ and $u^{\prime}_ i$ as expansions in terms of the old vectors $u_i, w_i$:
  \beq
  \label{newvectors1}
  w^\prime_i &=& w_i + \delta_\eta w_i ~,~~ \delta_\eta w_i = i \alpha_{ij}\; w_j + \beta_{ij} \; u_j~, \nonumber \\
    u^\prime_i &=& u_i + \delta_\eta u_i ~,~~\delta_\eta u_i = i \gamma_{ij}\; u_j + \kappa_{ij} \; w_j~,
  \eeq
  where $\alpha, \beta, \kappa, \gamma$ are assummed linear in $\eta$.
  Substituting (\ref{newvectors1}) into the orthonormality relations for the primed vectors, we immediately see that they require that $\alpha$ and $\gamma$ be hermitean matrices, while $\beta^\dagger = - \kappa$.
We now plug (\ref{newvectors1}) into (\ref{newvectors}), keeping terms to leading order in $\delta_\eta$, to find the equations determining the change of the vectors (a sum over repeated indices is assumed):
\beq
\label{newvectors2}
(1- \hat\gamma_5) \; \delta_\eta w_i &=& - (\gamma_5 \delta_\eta D)\;  w^i     ~ \leftrightarrow~  ~2 \beta_{ij} u_j = - (\gamma_5 \delta_\eta D) w_i  \nonumber \\
(1+ \hat\gamma_5) \; \delta_\eta u_i &=&  (\gamma_5 \delta_\eta D) \; u^i  ~ \leftrightarrow  ~ ~ 2 \kappa_{ij} w_j =  (\gamma_5 \delta_\eta D) u_i ~,
\eeq
showing that the hermitean matrices $\alpha$ and $\gamma$ are completely arbitrary, while $\beta$ and $\kappa$ are completely  determined by $\delta_\eta D$.  Finally, for the change of the basis vectors with a change of gauge background, $\delta_\eta w_i$ and $\delta_\eta u_i$ (it is easy to check that $\beta^\dagger = - \kappa $ for the explicit solution below), we find:
\beq
\label{newvectors3}  
  \delta_\eta w_i &=& i \alpha_{ij} w_j - {1 \over 2}\;  u_j (u_j^\dagger \cdot \gamma_5 \delta_\eta D \cdot w_i)~,\nonumber \\
\delta_\eta u_i &=& i \gamma_{ij} u_j + {1 \over 2} \; w_j (w_j^\dagger \cdot \gamma_5 \delta_\eta D \cdot u_i)~.
 \eeq
Eqns.~(\ref{newvectors3}) show that the change of the basis vectors with definite chirality  in the direction orthogonal to the same chirality subspace is completely determined by the change of the gauge background and that 
  the arbitrariness is in the freedom to make an unitary transformation in the given chirality subspace.
  
It is also clear from (\ref{newvectors3}) that the change of the basis vectors contributing to the change of measure and Jacobian, as in (\ref{J1}), can be written, by, e.g., changing the gauge background at a single link only and using linearity of $\alpha, \gamma$ in $\eta$, as follows: 
     \beq
  \label{vectorchange6}
  \sum_i (w_i^\dagger \cdot \delta_\eta w_i) &=& i \alpha_{ii} \equiv    - \sum_{x, \mu} \eta_{x, \mu} \;  j^{w}_{\mu, x}[U] ~,
  \nonumber \\
  \sum_i (u_i^\dagger \cdot \delta_\eta u_i) &=&  i \gamma_{ii} \equiv - \sum_{x, \mu} \eta_{x, \mu}  \; j^{u}_{\mu, x}[U]~. 
  \eeq
The ``currents" appearing in (\ref{vectorchange6})  are, generally, functionals of the gauge background as we have indicated  above; the left- and right-handed currents $j^u$ and $j^w$ can be  different.  We stress that the measure terms (\ref{vectorchange6}) are purely imaginary---it is precisely the $U$-dependence of the phase   of the chiral partition functions that is left ambiguous.

While the perturbative equations (\ref{newvectors}, \ref{newvectors1}) do not determine the currents (\ref{vectorchange6}), there are important restrictions imposed on them by global considerations \cite{Neuberger:1998xn}. These arise upon considering the second variation of the ``measure" terms (\ref{vectorchange6}), 
$\delta_\zeta \sum_i (w^\dagger_i \delta_\eta w_i) = \sum_i (\delta_\zeta w_i^\dagger\cdot \delta_\eta w_i) + ( w_i^\dagger \cdot \delta_\zeta \delta_\eta w_i) $,   in particular the ``curvature:"\footnote{That $f_{\zeta \eta}^w$ is indeed the sum of Berry curvatures for the positive ``energy" eigenstates $w_i$ of the ``Hamiltonian" $\hat\gamma_5$ depending on the parameters $[U]$ is explained in \cite{Neuberger:1998xn}.}
\beq
\label{curv1}
f_{\zeta \eta}^w \equiv \sum_i (\delta_\zeta w^\dagger_i \cdot \delta_\eta w_i) - (\delta_\eta w^\dagger_i \cdot \delta_\zeta w_i)~,
\eeq
which can be calculated upon substituting eqns.~(\ref{newvectors3}) for the variations $\delta_\eta w_i$, $\delta_\zeta w_i$ into (\ref{curv1}). One notices that $f_{\zeta\eta}^w$ is independent on the undetermined matrices $\alpha_{ij}$ and only depends on the variation of the basis vectors in the orthogonal subspace:
\beq
\label{curv2}
f_{\zeta\eta}^w &=& {1 \over 4}\; \sum_{i,j} \left( (w_i^\dagger \cdot \gamma_5 \delta_\zeta D \cdot u_j) (u_j^\dagger \cdot \gamma_5 \delta_\eta D \cdot w_i) - (\zeta \leftrightarrow \eta)\right) \nonumber \\
&=& {1\over 4}\;  {\rm Tr} \left( \hat{P}_+ \left[ \gamma_5 \delta_\zeta D, \gamma_5 \delta_\eta D \right]   \right) \nonumber \\
&=& {\rm Tr}\left( \hat{P}_+ \left[ \delta_\zeta \hat{P}_- , \delta_\eta \hat{P}_-\right] \right)~,
\eeq
where we used the relations between eigenvectors and projectors of Section \ref{gen1}. A similar relation is obtained for the negative chirality curvature:
\beq
\label{curv3}
f_{\zeta\eta}^u = {\rm Tr} \left( \hat{P}_- \left[ \delta_\zeta \hat{P}_+ , \delta_\eta \hat{P}_+\right] \right)~,
\eeq
obeying, of course, $f^u_{\zeta\eta} + f^w_{\zeta\eta} = 0$. 
The relations (\ref{curv2}, \ref{curv3}) show that the curvatures of the measure terms (\ref{vectorchange6}) are basis independent and imply that if the curvatures are nonvanishing, the the currents $j_\mu^w[U]$, $j_\mu^u[U]$---depending on the choice of phases (\ref{newvectors3}) of the $\hat\gamma_5$ eigenvectors $w_i$, $u_i$---can not be chosen to be independent on the background (in particular, they can not be taken to vanish).

Most importantly,     eqns.~(\ref{curv2}, \ref{curv3}) also imply that if perturbative anomalies do not separately  cancel among the light and mirror fermions, the measure terms $\sum_i (w^\dagger_i \cdot \delta_\eta w_i)$ can not be chosen to be smooth functions of the gauge field background (see Section \ref{anomalywilson} and \cite{Neuberger:1998xn},\cite{Luscher:1998du}).
This is because  the curvature defined above integrates, over some closed sub-manifolds in the gauge field configuration space, to quantized non-zero values.

The ``measure" terms determine the variations of the chiral partition functions
under changes of the gauge background---as we saw in the example of the vectorlike theory with the usual kinetic action, see
(\ref{z1}, \ref{deltaMplus}, \ref{deltaMminus}, \ref{deltaJ}).  We will prove  this for more general chiral partition functions in Section \ref{Yanwen'sTheorem}. The singular nature of the measure  implies that the separation of the partition function into chiral ``light" and ``mirror" components can not be a smooth function of the gauge background if  the light and mirror anomalies do not separately cancel and that the separation of the partition function is ill-defined. The explicit manifestation of the singularity in the 1-0 model will be considered in Sections \ref{anomalywilson}, \ref{measure}.

\subsection{On the variations of chiral partition functions}
\label{Yanwen'sTheorem}
  
Before considering anomaly cancellation and the smoothness of the measure, in this Section, we prove an important property on the variation of chiral partition functions.  It is a fairly straightforward proof but is also very general.  We find it quite useful, and in particular, Section~\ref{nirvana} contains an example of how such a general proof can lead to some strong conclusions.
  
Suppose $S[X^a_x,\; Y^{\dag b}_x,\; O^c_{xy}]$ is an arbitrary action.  Here $X^a_x$
and $Y^{\dag b}_x$ are some $d$-dimensional (``fundamental'' and ``anti-fundamental'')
vectors, $a$ and $b$ are ``flavor" indices, and $x=1, 2, \dots, d$ labels both spatial
and spinor indices.  $O^c$ are some additional operators the theory depends on, 
which we assume to be some $d\times d$ matrices. The action is said to be ``chiral'' in the following sense.  For each
flavor $X^a$ and $Y^{\dag b}$, there exist  projection operators $\hat P^a$ and $P^b$ respectively, all satisfying
$P^2=P$, such that:
\begin{equation}
\label{eq:def_chiral}
S[X^a_x,\; Y^{\dag b}_x,\; O^c_{xy}]=S[\hat P^a_{xy} X^a_y,\; Y^{\dag b}_yP^b_{yx},\; O^c_{xy}].
\end{equation}
A summation over all repeated lattice and spinor  indices ($x,y,z$) is understood, here and further in this section.
Given any action $\tilde S[X, Y^\dag, O]$ and some projection operators $\hat P$ and $P$, one can always ``build'' a chiral action by just defining $S[X, Y^\dag, O]=\tilde S[\hat{P} X, Y^\dag P, O]$.  The following property of a ``chiral'' action is essential for our 
discussions here.  Given any vector $u_x$ such that $\hat P^a_{xy} u_y=0$, one has:
\begin{equation}
\label{eq:prop_hori}
 \frac{\delta S}{\delta X^a_x} \; u_x=\frac{\delta S}{\delta(\hat P^a_{zy}X^a_y)}\; \hat P^a_{zx} u_x
=0,
\end{equation}
where equation \eqref{eq:def_chiral} is used in the second step. 
A similar property holds true for $\delta S/\delta Y^{\dag b}_x$ as well.

 We now proceed to explicitly construct the chiral partition functions by choosing two sets of orthonormal 
basis $w_i$ and $t_j$ such that:
\begin{gather}
\hat P w_i=w_i, \qquad P t_j=t_j\\
w^\dag_i w_j=\delta_{ij}, \qquad t^\dag_i t_j=\delta_{ij}~.
\end{gather}
Using chirality, we then set $X = \sum_i c_i w_i$, $Y^\dagger = \sum_i \bar{c}_i t_i^\dagger$
and define the partition function, specializing without loss of generality to the case of one flavor: 
\begin{equation}
\label{zs1}
Z[U]=\int\prod_i\; \ud c_i\prod_j\; \bar \ud  c_j\; e^{S\left[\sum\limits_i c_i w_i,\; \sum\limits_j \bar c_j t^\dag_j, \;O \right]}.
\end{equation}
We also note that no assumptions about the locality, (bi-)linearity, etc.,  of $S$ are made here; in particular, $S$ may be an effective action for the chiral fermions $X$ and $Y$ obtained after  integration over some other degrees of freedom;  its chirality (\ref{eq:def_chiral}) is the only important property for what follows. For example, $S$ could be the mirror-fermion effective action obtained after integrating over the random ($\kappa \rightarrow 0$) unitary higgs field in the ``1-0" model with mirror action given in (\ref{toymodel}, \ref{Skappa}). 

We now imagine that the projectors $\hat{P}$ as well as the operator(s) $O$ depend on some external fields $U$,  inducing external field dependence of $Z[U]$ as indicated in (\ref{zs1}); here we will assume that the projector $P$ can also depend on $U$, although in our application this will not be the case. We wish to compute the variation of $Z[U]$ under changes of the background field:
\begin{equation}
w_i\rightarrow w_i+\delta w_i, \qquad  t_i\rightarrow t_i+\delta t_i,
\end{equation}
and 
\begin{equation}
O \rightarrow O  + \delta O~.
\end{equation}
The variation of $S$ is given by:
\begin{equation}
\delta S=  \frac{\delta S}{\delta X_x}\sum_i c_i \; \delta w_{ix}+
\sum_j \bar c_j \; \delta t^\dag_{ix} \; \frac{\delta S}{\delta Y^\dag_x} +{\delta  S\over \delta O } \; \delta O~.
\end{equation}
The variation of the partition function is, therefore: \beq
\label{delz1}
\delta Z[U]&=& \int\prod_i \ud c_i \prod_j \ud \bar c_j \; e^S\;  \delta S \\
&=& Z[U]  \cdot \left[ \sum_i\left< \frac{\delta S}{\delta X_x} c_i\right> \delta w_{ix} 
+\sum_j \delta t^\dag_{jx} \left< \bar c_j\frac{\delta S}{\delta Y^\dag_x}\right> + \left< {\delta  S\over \delta O } \; \delta O\right>\right].\nonumber 
\eeq 
Here and below, ``$<\;>$'' denotes expectation values.

Now, the following identity of an arbitrary grassmann integral is easily verified
\begin{equation}
\label{eq:prop_grass}
\int \prod\limits_i  \ud c_i \; \frac{\delta F(c_1, c_2, \dots)}{\delta c_k} \; c_l=\delta_{kl}\; \int\prod\limits_i  \ud
c_i \; F(c_1, c_2,\dots). \end{equation}
 Here $F$ is an arbitrary function of
multiple grassmann numbers.  We are being very casual with
the ordering of grassmann numbers---this identity holds only 
if the ordering of the grassmann numbers are defined such that it's unchanged before and 
after the variation on $F$.  We will implicitly assume this rule in the following calculations.

It  is amusing that $\left<\frac{\delta S}{\delta X_x} \; c_i\right>$
and $\left<\bar c_j \; \frac{\delta S}{\delta Y_x}\right>$ can be computed without knowing the actual
form of $S$ at all,  essentially as a  direct consequence of identity
\eqref{eq:prop_grass} and the chirality of the action. We claim that:
\begin{equation}
\label{eq:frame_vev}
\left<\frac{\delta S}{\delta X_x} \; c_i\right>=w^\dag_{ix} \qquad \textrm{and}\qquad
\left<\bar c_j \;\frac{\delta S}{\delta Y_x}\right>=t_{jx}.
\end{equation}
To prove (\ref{eq:frame_vev}), one only needs to verify the inner products as:
\beq
  \left<\frac{\delta S}{\delta X_x} \; c_i\right> w_{jx}&=&\frac{1}{Z} \int \prod\limits_k \ud c_k \prod\limits_l \ud \bar c_l
\; \ e^S \; \frac{\delta S}{\delta X_x} \; c_i\; w_{jx} \nonumber \\
&=&\frac{1}{Z}\int \prod\limits_k \ud c_k \prod\limits_l \ud \bar{c}_l \;  \frac{\delta e^S}{\delta c_j}  \; c_i=\delta_{ij}
\eeq
with the help of identity \eqref{eq:prop_grass} in the last step.  For any other vector
$u_x$ that is  perpendicular to all the $w_i$'s one has: 
\begin{equation}
  \left<\frac{\delta S}{\delta X_x} c_i\right>  \; u_x=\frac{1}{Z} \int \prod\limits_k \ud c_k \prod\limits_l \ud \bar c_l
\; e^S \; \frac{\delta S}{\delta X_x} \; u_x \; c_i=0
\end{equation}
simply because $\frac{\delta S}{\delta X_x} \; u_x=0$, following from chirality of the action, eqn.~\eqref{eq:prop_hori}.
Similar properties are easily verified for 
$\left<\bar c_i\frac{\delta S}{\delta Y^\dag_x}\right>$.  Since the eigenvectors of $\hat{P}$ and the ones orthogonal to them form a complete set, these conditions are enough to 
conclude that equation \eqref{eq:frame_vev} holds true.

Therefore the variation (\ref{delz1}) of the partition function (\ref{zs1}) becomes, using (\ref{eq:frame_vev}):
\beq
\label{deltaZchiral}
\delta \log Z[U] =  \sum_i (w^\dag_i \cdot \delta w_i) +  \sum_i (\delta t^\dag_i \cdot t_i) + \left< {\delta  S\over \delta O } \; \delta O\right> ~.
\eeq
We thus showed that the factorization property of the variations of chiral actions alluded to after eqn.~(\ref{deltaMplus}) is  general---the variation of a chiral partition function always factorizes into a variation of the basis vectors plus a variation of the operators.

In the particular case when $\delta t_i=0$, $P=(1-\gamma_5)/2$, $\hat P=(1+\hat\gamma_5)/2$, and $Z[U]$ is the partition function of, say, the positive chirality fermion---defined by keeping the $c^+, \bar{c}^+$ integral in (\ref{z1}) only and equal to ${\rm det} \; (t_i^\dagger \cdot D\cdot w_j)$---it is clear that its variation, eqn.~(\ref{deltaMplus}), is reproduced by (\ref{deltaZchiral}).  

This theorem is also useful for determining how the chiral partition function transforms under any symmetry the original action $S[X, Y^\dag, O]$ happens to possess.  For example, suppose the action
respects the gauge symmetry, namely:
\begin{equation}
\label{eq:gauge_trans}
0=\delta_\omega S=\frac{\delta S}{\delta X} \delta_\omega X +\delta_\omega Y^\dag \frac{\delta S}{\delta Y^\dag}
+\frac{\delta S}{\delta O}\delta_\omega O,
\end{equation}
where:
\begin{equation}
\delta_\omega X=i \omega X,\qquad \delta_\omega Y=i \omega Y,\qquad \textrm{and}\quad \delta_\omega O=i [\,\omega, \,O\,]~,
\end{equation}
is the usual gauge transformation on the lattice.  Choosing
$P=(1+\gamma_5)/2$ and $\hat P=(1-\hat\gamma_5)/2$ and switching the notation
$w\rightarrow u$ and $t\rightarrow v$ for consistency, equations
\eqref{deltaZchiral} together with \eqref{eq:gauge_trans} immediately imply that under this transformation:
\begin{equation}
\label{eq:gauge_splitting}
\begin{split}
\delta_\omega \log Z=&\sum_i (u^\dag_i\cdot(\delta_\omega u_i-i\omega u_i))+i\sum_i (v^\dag\cdot\omega\, v)\\
=&\sum_i (u^\dag_i \cdot \delta_\omega u_i) -i \Tr\,\omega (\hat P -P)
=\sum_i (u^\dag_i\cdot \delta_\omega u_i)+\frac{i}{2}\Tr\,\omega\hat\gamma_5
\end{split}
\end{equation}
in agreement with equation \eqref{z03}.  This procedure applies to more general situations.
We will make further use of eqn.~(\ref{deltaZchiral}) in the following sections.

  \subsection{Anomaly cancellation and smoothness:  the  Wilson line background}
  \label{anomalywilson}
  
We now return to the issue of anomaly cancellation and the smoothness of the light-mirror split of the partition function.
It is well known that the existence of gauge anomaly in chiral
theories is deeply connected to the topology of the gauge field configuration space (for a discussion in the continuum, see \cite{AlvarezGaume:1984dr}, while for lattice overlap work, see \cite{Neuberger:1998xn}, \cite{Luscher:1999un}, \cite{Adams:2001jd}).
 On
the $2$-d square lattice with $U(1)$ gauge group, in a given topological  (flux) sector of admissible fields, this  space is a $N^2+2$ dimensional
torus times a contractible space \cite{Luscher:1998du} and the gauge anomaly prevents one from defining a smooth fermion measure in the
path-integral over this space.  The general properties of the gauge anomaly are 
discussed in \cite{Luscher:1998du, Luscher:1999un}, where it is proven that so long as the anomaly cancellation 
condition is satisfied a smooth measure exists.

In this Section,  following \cite{Neuberger:1998xn},  we focus on  two dimensional chiral 
theories with only homogeneous Wilson lines turned on, excluding all
other gauge field configurations (in sufficiently small volume, the Wilson lines give the leading contribution to the gauge path integral).  
The use of such a  simplified background is
that it allows us to  explicitly  construct  the fermion measure and literally see
where the singularities appear and how anomaly cancellation removes the 
difficulty.  

The two-dimensional theory is defined on a $N\times N$ lattice.  All the 
fields are endowed with periodic boundary conditions.
The gauge field configuration space in this sub-theory is completely
tractable.  We take the Wilson lines, denoted as $\mathbf h=(h_1, h_2)$, to be 
valued in the range $[0\, , 2\pi)$.  Physical quantities depending on them 
must be periodic functions with period $2\pi$.  This is the remnant
of the general gauge symmetry in this sub-theory. As a result, the variable $\mathbf h$ 
is valued on a two-torus defined by identifying the opposite sides of 
the square $[0\, , 2\pi]\times[0\, , 2\pi]$.  We shall refer to this torus 
as the $h$-torus, or $T^2_h$ in the following.

We would like to demonstrate how anomaly cancellation leads to a smooth measure in such a simplified
example.  First, we recall some known results of importance.  Consider the theory defined by 
a chiral action $S[X, Y^\dag, O]$ that satisfies the chirality property (\ref{eq:def_chiral}).  We assume that only the ``$\wedge$-ed''
projectors depend on the Wilson lines.  As is generally true in chiral theories 
on the lattice, the partition function is only defined with sets of basis vectors chosen 
for each projection operator.  Suppose they are chosen as:
\begin{gather}
\hat P_-(\mathbf h)\, u_i(\mathbf h)=u_i(\mathbf h),\qquad P_+ v_i=v_i~, \\
\hat P_+(\mathbf h)\, w_i(\mathbf h)=w_i(\mathbf h),\qquad P_+ t_i=t_i~,
\end{gather}
and define the partition function as usual:
\begin{equation}
Z(\mathbf h)=\int\prod_i \ud c_i^- \ud\bar c_i^+
\cdot\exp\left(S\left[\sum_i c_i^- u_i, \sum_j \bar c^+_j v_j^\dag, O\right]\right). 
\end{equation}
As we know, $Z(\mathbf h)$ defined in such a way depends on the choice of the basis.  In particular if
we had chosen $u_i'=U(\mathbf h)_{ij} u_j$, where $U$ is some $\mathbf h$
dependent unitary matrix, the partition function defined with the new basis defers from the
old one by a pure phase $\det U(\mathbf h)$.  

As explained in Section \ref{Yanwen'sTheorem}, given any chosen basis, the variation of the chiral partition function (\ref{deltaZchiral}) 
consists of two terms:
\begin{equation}
\delta \log Z=\sum_i (u^\dag_i \cdot \delta u_i) +\frac{1}{Z} \; \int\prod_i\ud c^-_i\ud\bar c^+_i
\,e^{S}\,\frac{\delta S}{\delta O}\delta O,
\end{equation}
where only the first term, referred to as the ``measure term," depends on the basis choice and it uniquely determines
the fermion measure \cite{Luscher:1999un}.  In what follows we denote it by:
\begin{equation}
\mathcal{J}_\mu=\sum_i (u^\dag_i \cdot \partial_\mu u_i),
\end{equation}
where $\partial_\mu\equiv \frac{\partial}{\partial h_\mu}$.  We will also refer to it as the ``connection" \cite{Neuberger:1998xn}
because the ``curvature'' associated to it defined as
$f_{\mu\nu}=\partial_\mu \mathcal J_\nu-\partial_\nu\mathcal J_\mu$ 
plays an important role in our discussion here.  As we have derived in Section \ref{changeofvectors}, 
see eqns.~(\ref{curv1}, \ref{curv2}), the curvature is  given by:
\begin{equation}
f_{\mu\nu}=\sum_i (\partial_\mu u^\dag_i\cdot\partial_\nu u_i )
-(\partial_\nu u^\dag_i \cdot \partial_\mu u_i)
=\Tr\left(\hat P_-\commut{\partial_\mu \hat P_-}{\partial_\nu \hat P_-}\right),
\end{equation}
and is independent on the basis choice.  Furthermore, its integral over the entire $T_h^2$ is not difficult to 
compute \cite{Neuberger:1998xn}.  In the case of a single charge-1 chiral fermion 
with projector $\hat P_-=\frac{1-\hat\gamma_5}{2}$ (as will be further discussed below in Section \ref{measure}) the integral of the curvature over the $h$-torus  turns out to be:
\begin{equation}
\label{eq:flux}
\int_{T_h^2} \,f_{\mu\nu}=-2\pi i~.
\end{equation}
Eqn.~(\ref{eq:flux}) 
 immediately implies that there does not exist an everywhere smooth ``connection'' $\mathcal J_\mu$
defined on the $h$-torus since $\partial T_h^2=\varnothing$.  Given any chosen basis of $u_i$'s, 
$\mathcal J_\mu$ must always be singular at least at some isolated points on $T^2_h$.
More generally, with multiple charged fermions, each fermion flavor of charge $q$ 
contributes to the curvature a term $\pm q^2 f_{\mu\nu}(q\mathbf h)$, where the 
sign depends on the chirality.  Even if the anomaly free condition is 
satisfied, namely $\sum q_+^2=\sum q_-^2$, the total curvature 
$f_{\mu\nu}^{\textrm{TOT}}=\sum_{q_-} q_-^2 f_{\mu\nu}(q_-\mathbf h)
-\sum_{q_+} q_+^2 f_{\mu\nu}(q_+\mathbf h)$
does not vanish since each term in the summation varies with $\mathbf h$ differently.   
Its integral over $T_h^2$, however, does vanish:
\begin{equation}
\int_{T_h^2}\,\sum_{q_\pm} f^{q_\pm}_{\mu\nu}=2\pi i\left(\sum q_+^2-\sum q_-^2\right)=0.
\end{equation}
It then allows for a smooth ``connection:''
\beq
\mathcal J_\mu=\sum_{i, q_-} (u^{\dag\, q_-}_i \cdot \partial_\mu u^{q_-}_i) +
\sum_{i, q_+} (w^{\dag\, q_+}_i \cdot \partial_\mu w^{q_+}_i),
\eeq
to be defined on $T^2_h$. Recall that the measure term is the basis-dependent variation of the chiral partition function. Hence, if the measure term can be chosen to be smooth, a smooth fermion measure exists, at least in this subspace of the gauge field space; see \cite{Luscher:1998du} for a general proof of the existence of smooth measure in anomaly-free $U(1)$ lattice gauge theories and \cite{Luscher:1999un} for arguments in the nonabelian case.

\subsubsection{Defining the measure of the anomaly free chiral partition function}
\label{measure}

We will demonstrate how such a smooth measure can be found in the anomaly free 
theories by first choosing an explicit set of basis vectors. 

Notice that the Wilson lines are a homogeneous background and the theory has 
a translational symmetry, hence it is   convenient to work with the 
momentum eigenstates.  On the lattice of size $N\times N$, momenta are discretized 
in units of $\frac{\pi}{N}$.  With the Wilson lines turned on, the momenta 
effectively become continuous. The Wilson line background shifts the  values of momenta in physical 
observables that depend on them by an amount of $\frac{\mathbf h}{2N}$, as we will see 
in the following.  With $h_{1,2}$ defined to  take their values in $[0\, ,2\pi)$, this shift exactly 
``fills in'' the gaps between the discrete momenta.  Therefore, momenta
shifted by the Wilson lines live on $2$-torus defined by identifying the opposite sides of 
the square $[0\, , \pi]\times[0\, , \pi]$.  We will refer to this torus (the Brillouin zone) as the 
momentum-torus, or $T^2_k$.

To proceed with the explicit construction and  choose a basis, we first define the following functions:
\begin{gather}
\label{abc}
a({\mathbf p})=1-\frac{1-2s_1^2-2s_2^2}{\sqrt{1+8 s_1^2 s_2^2}}, ~b({\mathbf p})=\frac{2 s_2 c_2}{\sqrt{1+8 s_1^2 s_2^2}}, ~
c({\mathbf p})=\frac{2 s_1 c_1}{\sqrt{1+8s_1^2 s_2^2}}, 
\end{gather}
where $s_{1,2}\equiv \sin p_{1,2}$ and $c_{1,2}\equiv \cos p_{1,2}$.  The ``momenta" 
$\mathbf p=(p_1, p_2)\in T^2_k$ live on the momentum-torus. The functions 
$a(\mathbf p)$, $b(\mathbf p)$ and $c(\mathbf p)$ 
just defined are periodic functions of period $\pi$ and therefore
smooth and well-defined everywhere on $T_k^2$.  In momentum space, the two-dimensional Neuberger-Dirac operator $D=1-\gamma_5 \hat\gamma_5$, see (\ref{hatgamma5}), with $\gamma_5 = \sigma_3$, 
 has the form: 
\begin{eqnarray}
\label{momentumGW}
D(\bf{p}) = \left( \begin{array}{c} a_{\bf{p}}   \cr - i c_{\bf{p}} + b_{\bf{p}} \end{array} \begin{array}{c}   -i c_{\bf{p}} - b_{\bf{p}} \cr  a_{\bf{p}} \end{array} \right)~,
\end{eqnarray}
and the GW relation $\hat\gamma_5^2=1$ is equivalent to $a^2_{\bf{p}} + b^2_{\bf{p}} + c^2_{\bf{p}} = 2 a_{\bf{p}}$.

Let us first focus on the case of a single fermion of charge $1$. A particularly simple choice 
of the basis vectors is given below \cite{Giedt:2007qg}.  For $\hat P_-$ eigenvectors, we choose:
\begin{equation}
\label{ubasis}
u_{\mathbf{k, h}}=\frac{1}{{\sqrt 2} N}\,e^{i 2 \mathbf k\cdot\mathbf x}\left[
\begin{array}{c}
\sqrt{a(\mathbf{k}+\frac{\mathbf{h}}{2N})}   \\\,\\
i\sqrt{2-a(\mathbf{k}+\frac{\mathbf{h}}{2N})}\;e^{i\varphi_{\mathbf{k}+\frac{\mathbf{h}}{2N}}} 
\end{array}\right],
\end{equation}
and for $\hat P_+$:
\begin{equation}
w_{\mathbf{k, h}}=\frac{1}{{\sqrt 2} N}\,e^{i 2  \mathbf k\cdot\mathbf x}\left[
\begin{array}{c}
i \sqrt{2-a(\mathbf{k}+\frac{\mathbf{h}}{2N})}  \\\,\\
 \sqrt{a(\mathbf{k}+\frac{\mathbf{h}}{2N})}\;e^{i\varphi_{\mathbf{k}+\frac{\mathbf{h}}{2N}}} 
\end{array}\right].
\end{equation}
 Here we defined the phase factor: 
\begin{equation}
\label{eiphi}
e^{i\varphi_{\mathbf p}}\equiv\frac{i b_{\mathbf{p}} +c_{\mathbf{p}}}
{\sqrt{b_{\mathbf{p}}^2+c_{\mathbf{p}}^2}},
\end{equation}
and the momenta:
\begin{equation}
\label{kvalues}
\mathbf k=\left(\frac{n\pi}{N}, \; \frac{m\pi}{N}\right),\qquad n,m=0,1,\dots,N-1.
\end{equation}
For the projectors  $P_\pm$ that are independent of the Wilson lines we simply choose: 
\begin{equation}
\label{tvbasis}
v^\dag_{\mathbf {k,h}}=\frac{1}{  N} e^{-i 2 \mathbf k\cdot \mathbf x}(\,1\quad 0\,),\qquad
t^\dag_{\mathbf {k,h}}=\frac{1}{  N} e^{-i  2 \mathbf k\cdot \mathbf x}(\,0\quad 1\,)~.
\end{equation}

In the chiral theory defined by an action $S[X, Y^\dag, O]=S[\hat P_- X, Y^\dag P_+, O]$,
only $u$'s (\ref{ubasis}) and $v$'s (\ref{tvbasis}) will be involved. Besides the ``wave-function'' $e^{i2\mathbf k\cdot\mathbf x}$ (which can be
varied by a gauge transformation), everything just defined indeed depends only 
on the combination $\mathbf p=\mathbf k+\frac{\mathbf h}{2N}$.  We find the following
picture sometimes helpful.  One can imagine that the discretized momenta $\mathbf k$ sit on the
sites of a $N\times N$ square lattice on the momentum torus $T^2_k$.  The effect of
the Wilson line $\mathbf h$ is to shift this lattice around $T^2_k$.  When $\mathbf h$ goes
one cycle around $T^2_h$, this lattice is shifted exactly by one unit cell and overlaps 
with the original.

Notice that the function $e^{i\varphi_\mathbf p}$ of (\ref{eiphi}) is ill-defined\footnote{The zero gauge background vectors used to split the partition function in  \cite{Giedt:2007qg} have a discontinuity at these values of momenta.}  at 
$\mathbf p=\mathbf k+\mathbf h/(2N)=(0, 0), (\frac{\pi}{2}$, $\frac{\pi}{2})$, $(0,\frac{\pi}{2})$, and $(\frac{\pi}{2},0)$.
 Given
that $\mathbf k$ is discretized (\ref{kvalues}), these points are only (for unit values of the charge) reached by certain modes of $\mathbf k$
when $\mathbf h=(0, 0) \mod 2\pi$.  As a consequence, the ``connection:"
\begin{equation}
\label{current1}
\mathcal J_\mu(\mathbf h)=\sum_{\mathbf k} ( u^\dag_{\mathbf {k, h}} \cdot \partial_\mu u_{\mathbf {k, h}})
=  \frac{i}{2}\sum_{\mathbf k} \left( 2 - a(\mathbf k+\frac{\mathbf h}{2N})\right) \;
\partial_\mu\varphi_{\mathbf k+\frac{\mathbf h}{2N}}~,
\end{equation}
as a vector field defined on $T^2_h$, can be singular only at $\mathbf h=(0,\,0)$ and
is perfectly smooth everywhere else.
The singularity at $\mathbf h=(0,\,0)$ is expected since we already know from (\ref{eq:flux}) as well
as \cite{Neuberger:1998xn},  that the ``curvature'' associated to ``connection'' $\mathcal J_\mu$
integrates over the entire torus to $2\pi i$. (The curvature $f_{\mu\nu}$ and its integral can also be
explicitly computed from (\ref{current1}), using (\ref{eiphi}, \ref{abc}).) Imagine on $T_h^2$
we draw a small circular disc $D$ of radius $r$ centered at $\mathbf h=(0,\,0)$; then, by Stoke's theorem:
\begin{equation}
\int_{\partial D}\mathcal J_\mu=-\int_{T^2_h-D} f_{\mu\nu}.
\end{equation}
Since $f_{\mu\nu}$ is smooth and finite everywhere on $T^2_h$, in the limit $r\rightarrow 0$,
\begin{equation}
\lim_{r\rightarrow 0}\int_{\partial D} \mathcal J_\mu=-\int_{T_h^2} f_{\mu\nu}=2\pi i.
\end{equation}
Together with symmetry considerations, we are led to the conclusion that
$\mathcal J_\mu$ diverges as:
\begin{equation}
\mathcal J_\mu(\mathbf h\rightarrow 0) \approx \frac{1}{r}\hat\theta_\mu.
\end{equation}
Here $r=\sqrt{h_1^2+h_2^2}$ and $\hat\theta_\mu$ denotes the unit vector tangential 
to $\partial D$.  Exactly   this $\frac{1}{r}$-singularity is what prevents us 
from defining the measure smoothly.  If we think of
$\mathcal J_\mu$ as a vector field defined on $T_h^2$, this singularity
appears as a divergent vortex around the singular point. As explained above, 
 this is the only singularity of $\mathcal J_\mu$, and if removed, 
$\mathcal J_\mu$ is smooth.

\begin{figure}[!htb]
\begin{center}
\includegraphics[width=0.9\textwidth]{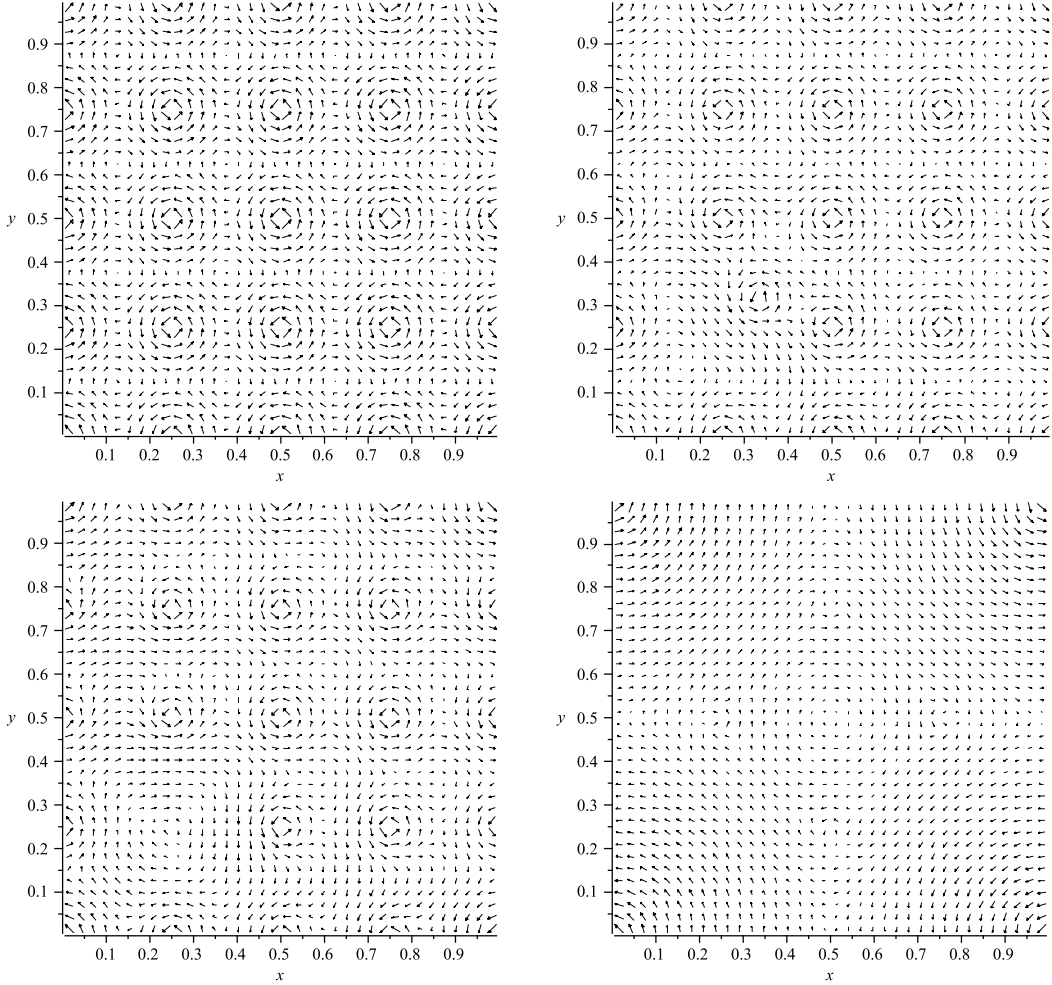}
\caption{\label{fig:vortices4} \small First panel: the 16 singularities of $\mathcal J_\mu^4$, each
vortex has a divergence at the center.
Second: one  vortex is slightly shifted.
Third: one   vortex is moved all the way to $\mathbf h=(0,\,0)$
so that two singularities coincide there; Fourth: all the vortices are shifted to the
corner (the strength of the singularity in the fourth panel is scaled to that of a single vortex).  Axes $(x,y)=(h_1/2\pi, h_2/2\pi)$. }
\end{center}
\end{figure}

Let us generalize these results to fermions of charge $q$
in a simple manner.  Just replace all the $\mathbf h$'s in the expressions 
for $u_{\mathbf {k, h}}$ and $w_{\mathbf {k, h}}$ by $q \mathbf h$.  
The ``measure term'' is modified to:
\begin{equation}
\mathcal J^q_\mu(\mathbf h)=\sum_{\mathbf k} (u^{\dag}_{\mathbf {k}, q \mathbf h} \cdot
\partial_\mu u_{\mathbf {k}, q\cdot\mathbf h}) = q \; \mathcal J^1_\mu(q\mathbf h).
\end{equation}
Near every singular point of $\mathcal J^q_\mu$, the properties just discussed 
above continue to hold.  For example, near the
point $\mathbf h=(0,\,0)$, where the measure term $\mathcal J^q_\mu$ diverges for any value of $q$, we have:
\begin{equation}
\mathcal J^q_\mu(\mathbf h\rightarrow 0)=q \mathcal J^1_\mu(q \mathbf h\rightarrow 0)
\approx q\cdot \frac{1}{q r}\hat\theta=\frac{1}{r}\hat\theta.
\end{equation}
Hence, its line integral around the singularity is still $2\pi i$.
However, as $\mathcal J^q_\mu$ depends on $\mathbf h$ through $q \mathbf h$,
by the periodic properties of $\mathcal J^1_\mu$, the number of locations where
$\mathcal J^q_\mu$ diverges increases to  $q^2$.  Indeed, instead of 
having singularity only at $\mathbf h=(0,\,0)$, the same type of singularity must 
repeat itself at every point where $\mathbf h=(\frac{2n\pi}{q},\,\frac{2m\pi}{q}), n,m=0,1,\dots, q-1$, 
exactly the right amount  to account for the integral of 
$f^q_{\mu\nu}$ over $T_h^2$ that scales as $q^2$. Fig.~\ref{fig:vortices4}
illustrates the $16$ singularities of $\mathcal J^4_\mu$ on $T^2_h$ given by
a chiral fermion of charge-$4$.  Each vortex indicates a $\frac{1}{r}$-type divergence at 
its center.  The four corners and the two opposite sides are to be identified.

Evidently, the singularities we just discovered are the manifestation of the topological
obstruction that prevents one from defining a smooth measure for the anomalous chiral theory  \cite{Neuberger:1998xn, Luscher:1998du}.  
We now focus on the anomaly free case, namely when $\sum q_+^2=\sum q_-^2$ is satisfied. 
Obviously fermions with opposite chirality produce vortices with opposite signs,
and if they sit on top of each other, they cancel.  The anomaly-free condition
guarantees that there are always equal number of $+$ and $-$ vortices, giving
a nice understanding of the fact that the integral of $f^{\textrm{total}}_{\mu\nu}$ 
over the $T_h^2$ vanishes.  For the purpose of defining a smooth measure though, 
this is not sufficient, since normally $+$ and $-$ vortices do not just sit on top of each other.
With the current choice of basis, the singularities produced by each charge-$q$ chiral fermion are 
  distributed on $T_h^2$ with equal separations in both directions.  This regularity 
is very helpful for the counting, but not for the smoothness of the measure.
In the ``$345$'' model for an example, only the vortices at $\mathbf h=(0,\,0)$ overlap and all 
the rest miss each other.  As the consequence,
$\mathcal J_\mu=\mathcal J_\mu^3+\mathcal J_\mu^4-\mathcal J_\mu^5$ diverges
at many places.

To find a smooth measure term $\mathcal J_\mu$, one only needs to utilize the freedom of basis choice.
With the translational symmetry of the Wilson line background to be respected,
we are left with only the option to multiply the vectors by some 
$\mathbf h$ dependent phases.  This turns out to be sufficient.  If we choose to replace the basis vectors (\ref{ubasis}) 
$u_{\mathbf{k}, q \mathbf{h}}\rightarrow u_{\mathbf{k}, q \mathbf{h}} e^{ i\sigma_{\mathbf{k},q \mathbf{h}} }$, 
the new measure reads:
\begin{equation}
\mathcal J_{\mu}\rightarrow \mathcal J_{\mu}+i\sum_{\mathbf k, q}\partial_\mu \sigma_{\mathbf{k},q \mathbf{h}}.
\end{equation}
The additional term $\partial_\mu\left(\sum \sigma\right)$ is a total derivative of 
some function defined on the $h$-torus.  If it has no singularities of its own, it certainly
does nothing interesting. If it has the same vortex-type singularities as those
found in the measure, positive and negative vortices must appear in pairs, because the curl of $\partial_\mu\sigma$ vanishes. 
Therefore, one can imagine designing such a $\sigma_q$, so that $\partial_\mu\sigma_q$ has at least a pair 
of $+$ and $-$ divergent vortices and one of the them coincides with
one of the singularities of $\mathcal J^q_\mu$ but with an 
opposite sign.  This will cancel that particular singularity of $\mathcal J^q_\mu$ 
at the position where it was, but will create  it  elsewhere.  The net effect is that singularities can be moved
at will through such a manipulation.  The second panel of Fig.~\ref{fig:vortices4} demonstrates
a particular choice of $\sigma$ that slightly shifts one of the singularities 
emerging in $\mathcal J^4_\mu$. 

\begin{figure}[!htb]
\begin{center}
\includegraphics[width=0.6\textwidth]{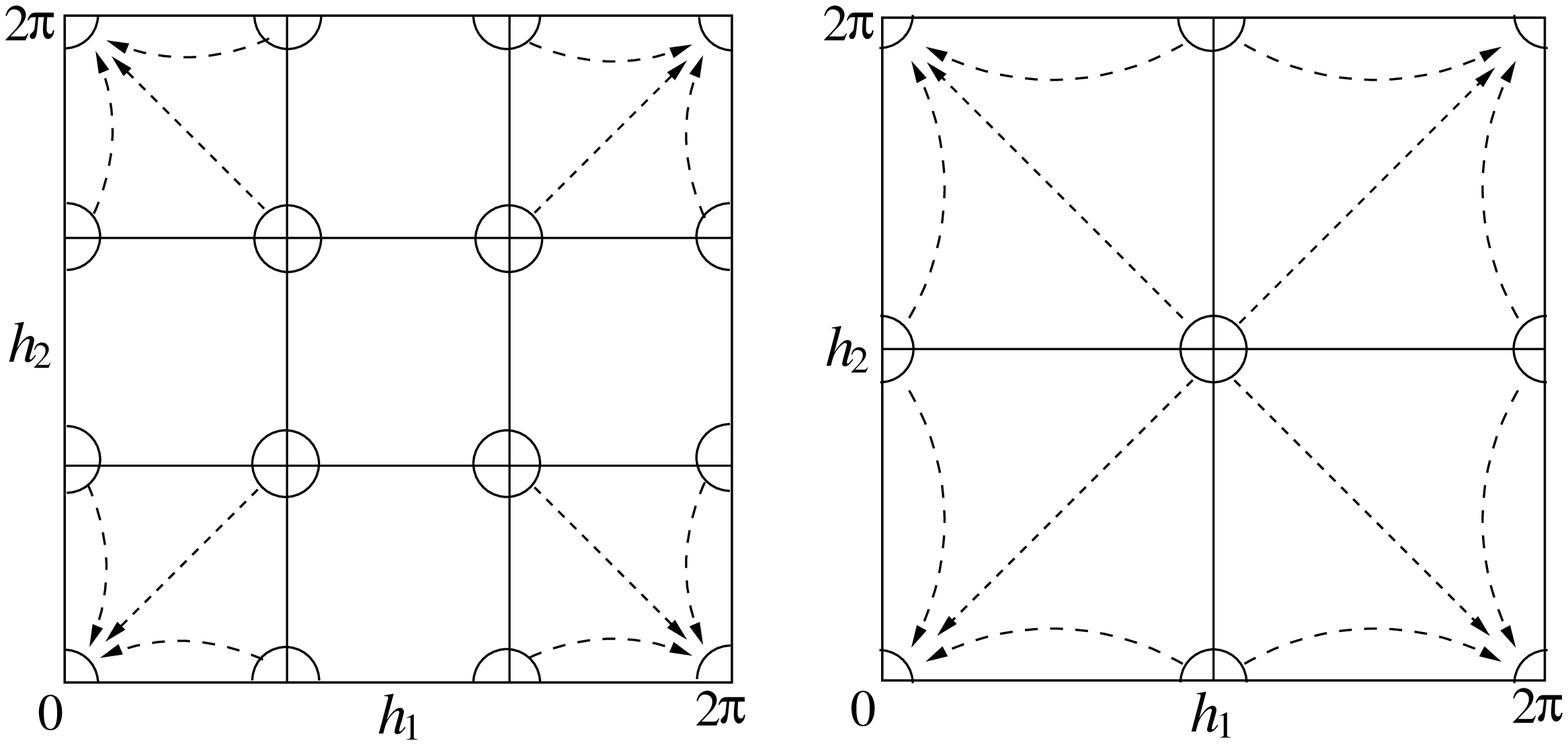}
\caption{\label{fig:moving}\small A pictorial illustration for moving the singularities
of $\mathcal J^3_\mu$ and $\mathcal J^2_\mu$ respectively.  Each circle represents
a divergent vortex and the arrows denote how they should be shifted. The second panel represents the operation described by eqn.~(\ref{abcd}).}
\end{center}
\end{figure}

We can now envision how a smooth measure can be defined in the anomaly-free case.
Simply design the function $\sigma_q$ such that all the singularities of $\mathcal J^q_\mu$
are shifted to a common place so that they can be cancelled by the singularities of appropriate opposite-chirality fermions.  A simple way of doing so is to move every vortex 
toward $\mathbf h=(0,\,0)$.  During the procedure, one might wish to preserve the lattice
rotational symmetry.  Such a constraint can be obeyed by moving the singularities 
in a $Z_4$ symmetrical way, as pictorially illustrated by Fig.~\ref{fig:moving} 
for moving the singularities of $\mathcal J^3_\mu$ and $\mathcal J^2_\mu$ respectively.

An explicit expression for $\sigma$ that realizes the manipulations illustrated
in Fig.~\ref{fig:moving} can be constructed by first defining 
$T(x)=\tan\left(\frac{x-\pi}{2}\right),$
and for the charge-2 term $\mathcal J^2_\mu$ as an example, choose $\sigma$ to be ($e^{i \sigma}$ is to be applied on only  one of the basis vectors $u_{\mathbf{k}, q \mathbf{h}}$):
{\small
\begin{equation}
\begin{split}
\label{abcd}
\sigma(h_1, h_2)=&\frac{1}{4}\left[\tan^{-1}\frac{T(h_2)}{T(h_1-\pi)-T(h_1)}
-\tan^{-1}\frac{T(2\pi-h_2)}{T(h_1-\pi)-T(h_1)} \right.\\
&\left.-\tan^{-1}\frac{T(h_2)}{T(\pi-h_1)-T(2\pi-h_1)}
+\tan^{-1}\frac{T(2\pi-h_2)}{T(\pi-h_1)-T(2\pi-h_1)}\right]\\
+&\frac{1}{4}\left[-\tan^{-1}\frac{T(h_1)}{T(h_2-\pi)-T(h_2)}
+\tan^{-1}\frac{T(h_1)}{T(\pi-h_2)-T(2\pi-h_2)}\right.\\
&\left.+\tan^{-1}\frac{T(2\pi-h_1)}{T(h_2-\pi)-T(h_2)}
-\tan^{-1}\frac{T(2\pi-h_1)}{T(\pi-h_2)-T(2\pi-h_2)}\right]\\
-&\frac{1}{2}\tan^{-1}\frac{T(h_2)}{T(h_1)}
+\frac{1}{2}\tan^{-1}\frac{T(h_1)}{T(h_2)}\;.
\end{split}
\end{equation}}
{\flushleft A}s long as
all the $\frac{1}{r}$-type singularities are cancelled, the new measure
$\mathcal J_\mu+i\partial_\mu\left(\sum\sigma\right)$ is smooth everywhere.  

 \section{Resolution of the paradox and smoothness of the mirror partition function in anomaly-free case}
  \label{nirvana}
  The resolution of the paradox of Section \ref{paradox} should be evident by now. The basis vectors used to split the light and mirror partition functions in the ``1-0" model have a discontinuity at the four special points in momentum space (see footnote before eqn.~(\ref{current1})). This discontinuity of the basis vectors  causes  them and the measure to be singular already when only the Wilson line sector of gauge field space is considered (at $h_\mu = 0$).  This   singularity  can not be removed by redefining their phases and is related to the nonvanishing chiral anomaly in the mirror  and light sectors. Thus, while the results of \cite{Giedt:2007qg} hold at $U=1$, the mirror partition function, generating functional, and spectrum are not smooth functions of the gauge background, and the trivial background results can not be used to infer anything about the spectrum when gauge background fluctuations are included.
 
 More interestingly than resolving the paradox, however, our results from Section \ref{Yanwen'sTheorem}, combined with those of ref.~\cite{Luscher:1998du} 
(proving that the smooth measure exists iff the anomalies cancel) imply that the mirror generating functional of a Yukawa-Higgs-GW model will be a smooth function of the gauge background whenever the mirror and light sectors are separately anomaly free (the proof below holds in finite volume).
  Most generally, we wish to prove that,  given the ``measure term":
\begin{equation}
\mathcal{J}_\mu=\sum_i (u^\dag_i \cdot \delta_\mu u_i)~,
\end{equation}
is smooth (here $\delta_\mu$ indicates variations in all possible directions in gauge field configuration space)  
the partition function defined by
\begin{equation}
\label{eq:def_chiral_partition}
Z=\int\ud \bar c \; \ud c \, \tilde S[c_i u_i, \,\bar c_i v_i^\dag, \,O]
\end{equation}
is always smooth so long as the operator(s) $O$  are smooth functions of the gauge
field and $\tilde S$ is a smooth functional of the operators.   We assume that
$u_i$ form an orthonormal basis of the $+1$-eigenspace of the operator $\hat P_-$, and that $\tilde S$ satisfies the usual chiral property: $\tilde S[X, Y^\dag, O]=S[\hat P_- X, Y^\dag, O]$ (we ignored the ``chiral property'' regarding the $Y$-fields here as those are assumed to be
gauge field independent).  Instead of $e^S$, we wrote $\tilde S$ to avoid
dealing with possible logarithms in the proof which might cause unnecessary doubts. We have in mind that $\tilde S$ is, for example, given by the mirror fermion action averaged over the random unitary Higgs field(s) $\phi$, the result of such averaging is a sum of multi-fermion terms, which are polynomials in terms of $c_i u_i,$ and $\bar c_i v_i^\dag$. 

We say that $Z$ is a ``chiral partition function'' in the general sense if it is defined by equation 
\eqref{eq:def_chiral_partition} with some $\tilde S$ that satisfies all the properties mentioned.  We remind
the reader that although the vectors $u_i$ might be ill-defined at certain isolated points in the
gauge field configuration space, they never diverge---they can not simply because they are unit 
vectors (see Section \ref{measure} for example where we constructed 
the smooth measure $\mathcal J_\mu$ using them in the Wilson-line subspace with anomaly free contents).
As a consequence, any ``chiral partition function'', being a grassmann integral defined with a smooth 
$\tilde S$, is always a finite function of the gauge field, even when evaluated infinitely close to 
the points where the basis vectors are ill-defined.  More precisely stated, within any compact 
region in the gauge configuration space, the absolute value of any ``chiral partition function'' is bounded 
from above (with a fixed lattice size). We will loosely use the word ``finite'' in the following to describe this property.

The proof for the smoothness of $Z$ is then really simple. One only needs to first notice that the variation of the action due to the variation of the operator $O$: 
\begin{equation} 
\tilde S'[X, Y, O]\equiv\frac{\delta \tilde S[X, Y, O]}{\delta O} \; \delta O~, 
\end{equation} 
is usually no longer chiral. However, if one defines: 
\begin{equation} 
\tilde S^{(1)}[X, Y, O]\equiv \tilde S'[\hat P_- X, Y, O]=\left.\frac{\delta \tilde S[X', Y, O]} 
{\delta O} \; \delta O \right|_{X'=\hat P_- X}, 
\end{equation} 
it is manifestly chiral. It is easily verified that:
\begin{displaymath}
\int \ud\bar c\ud c\; \tilde S^{(1)}[c_i u_i, \bar c_i v_i^\dag, O]= 
\int \ud\bar c\ud c\; \tilde S'[c_i u_i, \bar c_i v_i^\dag, O]. 
\end{displaymath}
Furthermore $S^{(1)}$ is a smooth functional of $O$ since the original action $S$ and the operators 
are smooth as we assumed. Therefore, whenever $Z$ is a ``chiral partition function,'' $Z^\prime$, defined by: 
\begin{equation} 
Z'\equiv\int\ud \bar c\ud c \,\tilde S^{(1)}[c_i u_i, \,\bar c_i v_i^\dag, \,O]= 
\int\ud \bar c\ud c \,\frac{\delta \tilde S[c_i u^i, \bar c_i v_i^\dag, O]}{\delta O}\delta O, 
\end{equation} 
is also ``chiral'' and thus finite as well.  By the ``splitting-theorem'' of section \ref{Yanwen'sTheorem},
we have
\begin{equation}
\label{eq:splitting}
\delta_\mu Z=Z\mathcal J_\mu+Z'.
\end{equation}
Given that $Z$, $Z^\prime$ and $\mathcal J_\mu$ are all finite, we immediately know the first variation of $Z$, smooth or not, is at least finite.

We are now ready to claim, by applying the same logic iteratively, that given the assumptions listed 
above (smoothness of ${\mathcal J}_\mu$, $\tilde S$ and $O$), 
any ``chiral partition function'' $Z$ (\ref{eq:def_chiral_partition}) is smooth. This is because for
$\forall n\in \mathbb Z$, the $n$-th derivative $Z^{(n)}$ can always be expressed as 
a polynomial in terms of some other ``chiral partition functions'' (which are always finite)
and some smooth functions (the measure term $ \mathcal{J}_\mu$ and its variations).  This is certainly true 
when $n=1$ as equation \eqref{eq:splitting} says.
Assuming the hypothesis holds true for some value of $n$, to prove
that it remains true for $n+1$ is almost trivial.  Just apply the agove procedure on
each ``chiral partition function'' appearing in the polynomial and recall that the derivative
of any smooth function is still smooth.  Hence, by induction this is true for any $n$. 
Because any ``chiral partition function'' is finite, so is $Z^{(n)}$.  
Therefore $Z$ is smooth.  Again, ``finiteness'' here means the function is
bounded within any compact region in 
the gauge configuration space.\footnote{To be mathematically precise, we remind the reader that
because the basis vectors are ill-defined at some
isolated points in the gauge configuration space, the partition function $Z$, defined by
\eqref{eq:def_chiral_partition}, rigorously speaking is only defined everywhere away from those places.
However, by showing the finiteness of all the derivatives evaluated infinitely close
to those places, we have proved that those points are  removable
singularities of $Z$, namely near any one of those points $x_0$, $\lim_{x\rightarrow x_0} Z(x)$ 
exists, and is well-defined and finite.  As long as we define $Z(x_0)=\lim_{x\rightarrow x_0}Z(x)$, $Z$ is a
smooth function on the entire gauge configuration space.}

   Thus, the smoothness of the mirror partition function (and generating functional, with source terms for the mirror fields added)  implies that an analytic or numerical result  that would indicate the decoupling of the mirror sector (at strong Yukawa coupling, say, as in \cite{Giedt:2007qg}) at vanishing gauge background would be expected to hold at least for ``nearby" gauge backgrounds, e.g., in perturbation theory with respect to the gauge coupling. We think that this  result clearly encourages further study of mirror-sector Yukawa-Higgs dynamics in anomaly free models. 
  
\bigskip

\bigskip

\section*{Acknowledgements}

We thank Joel Giedt, Maarten Golterman, and Yigal Shamir for discussions and comments. 
We also acknowledge support by the National Science and Engineering Council of Canada (NSERC).

\section{Addendum:} 
In the first paragraph of Section 5, we discussed the resolution to the paradox posed by the results of \cite{Giedt:2007qg}, in 
the situation that a dynamical gauge field is turned on. We argued that with a 
dynamical gauge field, the splitting of the fermions into different chiralities used in 
\cite{Giedt:2007qg} is mathematically inconsistent, and that the numerical evidence found in \cite{Giedt:2007qg}---using only the singular mirror partition function and indicating a complete decoupling of the mirror sector---can not be used to 
infer properties of the full theory with dynamical gauge fields. 
While this claim is plausible, it does not completely explain away the paradox, as it leaves still unresolved questions if the gauge field is treated as an external background. In the interest of completeness we wish to briefly explain these questions here. We hope to return to their detailed study in the near future. 

Since the obstruction to smoothness of the light-mirror split of the partition function is topological, given any gauge field background one can always 
find a choice of splitting that is smooth locally, in a small neighborhood in gauge field 
configuration space near the given point. If one treats this gauge field as a 
fixed external background only, some questions still remain. We point out an interesting observation 
here. Suppose 
$S=[X, Y^\dag, O(A_\mu(x)]$ is any gauge invariant chiral action which satisfies: 
\begin{equation} 
S[X, Y^\dag, O(A)]=S[\hat P(A) X, Y^\dag, O(A)]=S[X, Y^\dag P, O(A)]. 
\end{equation} 
Here $\hat P(A)=\frac{1-\hat \gamma_5(A)}{2}$ and $P=\frac{1+\gamma_5}{2}$, $A_\mu(x)$ is the 
external gauge field and $O$ represents all the operators appearing in the definition of $S$ 
and typically depend on the gauge field $A$. With some orthonormal bases $u_i$ 
and $v_i$, which are the appropriate eigenvectors of $\hat P$ and $P$ respectively, with $u_i$ chosen to be smooth 
with respect to $A$ near, say, $A=0$, the chiral partition function is defined as: 
\begin{equation} 
Z[A]=\int \ud c_i \ud \bar c_i \exp{S[c_i u_i, \bar c_i v_i^\dag, O(A)]}. 
\end{equation} 
We would like to calculate the polarization operator of $A_\mu$ at zero gauge field background, 
given by: 
\begin{equation} 
\Pi_{\mu\nu}(x,y)\equiv\left. \frac{\delta^2 \ln Z[A]}{\delta A_\mu(x) \delta A_{\mu}(y)}\right|_{A=0}. 
\end{equation} 
Using the theorem of Section 4.2, it is easily verified that $\Pi_{\mu\nu}$ splits 
into two parts.\footnote{Notice that while computing the higher derivatives of $\ln Z$, one must follow 
the procedure outlined in Section 5.} 
The first part comes from the variations of the ``measure term,'' due to the 
variation of $\ln Z$ caused by varying the eigenvectors $u_i$ with respect to the gauge field. This 
part is not interesting to us. In particular, if we embed this chiral theory into any vector-like 
theory, this part is cancelled by the contributions from the fermions with opposite chirality and that of 
the Jacobian. The second part of $\Pi_{\mu\nu}(x,y)$ appears while one varies $\ln Z$ by varying 
the operators $O(A)$ with respect to the gauge field. This piece is physically 
more interesting. In the following discussion, we focus only on this ``reduced'' polarization: 
\begin{equation} 
\label{eq:fermion_loop} 
\Pi'_{\mu\nu}(x,y)\equiv\left. \frac{\delta'^2 \ln Z[A]}{\delta' A_\mu(x) \delta' A_{\mu}(y)} 
\right|_{A=0}, 
\end{equation} 
where $\delta'$ means variations with respect to $A$ while keeping $u_i$ fixed 
as constant vectors. Clearly $\Pi'_{\mu\nu}(x,y)$ can be expressed as some complicated 
fermion 2-point correlators in this theory and $\Pi'_{\mu\nu}=\Pi'_{\nu\mu}$. While 
evaluated on a translationally symmetrical background (e.g., $A=0$), it depends on $|x-y|$ only. 

The divergence of this reduced 2-point function is easily calculated, since: 
\begin{equation} 
\begin{split} 
\sum_{\mu} \nabla^\ast_{\mu y} \frac{\delta'\ln Z[A]}{\delta' A_{\mu}(y)\delta' A_{\nu}(z)} 
=&-\frac{\delta}{\delta \omega(y)}{\sum_{\mu, x}\frac{\delta'\ln Z[A]}{\delta' A_{\mu}(x)\delta' A_{\nu}(z)}} 
\; \nabla_{\mu x} \omega(x)\\ 
=&-\frac{\delta'}{\delta' A_\nu(z)}\frac{\delta}{\delta \omega(x)}\sum_{\mu, y}\frac{\delta'\ln Z[A]} 
{\delta' A_{\mu}(y)} 
\nabla_{\mu y}\omega(y)\\ 
=&\frac{\delta'}{\delta' A_\nu(z)}\frac{\delta}{\delta \omega(x)}\delta'_\omega \ln Z[A]. 
\end{split} 
\end{equation} 
Here $\delta'_\omega \ln Z[A]$ is the variation of $\ln Z[A]$ under the arbitrary gauge variation 
$A_\mu(x)\rightarrow A_\mu(x)-\nabla_\mu \omega(x)$, while keeping the basis vectors $u_i$ fixed. 

We have assumed that $S$ is gauge invariant. Given this assumption, by equation \eqref{eq:gauge_splitting}, 
$\delta'_\omega \ln Z[A]$ is known to be exactly\footnote{As usual, see Section 3, this can be derived by first embedding the chiral 
theory into a vector-like theory defined by $S_{\textrm{full}}=S[X, Y^\dag, O]+Y^\dag(1-P)D(1-\hat P) X$ and 
then inferring the result from gauge invariance of the full theory.} 
$\frac{i}{2}\Tr\omega \hat \gamma_5$, 
completely independent to the details of $S$. It vanishes if and only if the anomaly cancellation 
condition is satisfied. Therefore in any anomalous chiral theory defined with projection operators $P$ 
and $\hat P$ whose classical action is gauge invariant, there exists a fermion 2-point 
correlation function defined by \eqref{eq:fermion_loop}, whose divergence is purely imaginary and 
proportional to $\delta \tr \hat\gamma_{5 xx}/{\delta A_\nu(y)}$. Even though this expression is 
local, it is known that it is not the divergence of a local expression. 
Therefore, the fermion correlator, as part of the gauge field polarization operator, must contain a nonlocal contribution. The physical interpretation of this fact and its manifestation 
in the 1-0 model requires further studies. In particular, it will be interesting to see how it shows 
up in the numerical simulations. We hope to report on this subject in follow up work soon.

\end{document}